\PassOptionsToPackage{unicode}{hyperref}
\PassOptionsToPackage{hyphens}{url}
\documentclass[
  12pt,
]{article}
\usepackage{lmodern}
\usepackage{amssymb,amsmath}
\usepackage{ifxetex,ifluatex}
\usepackage{arydshln}
\usepackage{mathtools}
\ifnum 0\ifxetex 1\fi\ifluatex 1\fi=0 
  \usepackage[T1]{fontenc}
  \usepackage[utf8]{inputenc}
  \usepackage{textcomp} 
\else 
  \usepackage{unicode-math}
  \defaultfontfeatures{Scale=MatchLowercase}
  \defaultfontfeatures[\rmfamily]{Ligatures=TeX,Scale=1}
\fi
\IfFileExists{upquote.sty}{\usepackage{upquote}}{}
\IfFileExists{microtype.sty}{
  \usepackage[]{microtype}
  \UseMicrotypeSet[protrusion]{basicmath} 
}{}
\makeatletter
\@ifundefined{KOMAClassName}{
  \IfFileExists{parskip.sty}{%
    \usepackage{parskip}
  }{
    \setlength{\parindent}{0pt}
    \setlength{\parskip}{6pt plus 2pt minus 1pt}}
}{
  \KOMAoptions{parskip=half}}
\makeatother
\usepackage{xcolor}
\IfFileExists{xurl.sty}{\usepackage{xurl}}{} 
\IfFileExists{bookmark.sty}{\usepackage{bookmark}}{\usepackage{hyperref}}
\hypersetup{
            pdftitle={Semiparametric Regression for Dual Population Mortality},
            pdfauthor={Gary Venter and Şule Şahin},
            pdfborder={0 0 0},
            breaklinks=true}
\usepackage[margin=1in]{geometry}
\usepackage{color}
\usepackage{fancyvrb}

\DefineVerbatimEnvironment{Highlighting}{Verbatim}{commandchars=\\\{\}}
\usepackage{framed}
\definecolor{shadecolor}{RGB}{248,248,248}
\newenvironment{Shaded}{\begin{snugshade}}{\end{snugshade}}

\newcommand{\CommentTok}[1]{\textcolor[rgb]{0.56,0.35,0.01}{\textit{#1}}}

\newcommand{\ControlFlowTok}[1]{\textcolor[rgb]{0.13,0.29,0.53}{\textbf{#1}}}
\newcommand{\DataTypeTok}[1]{\textcolor[rgb]{0.13,0.29,0.53}{#1}}
\newcommand{\DecValTok}[1]{\textcolor[rgb]{0.00,0.00,0.81}{#1}}

\newcommand{\ErrorTok}[1]{\textcolor[rgb]{0.64,0.00,0.00}{\textbf{#1}}}

\newcommand{\FloatTok}[1]{\textcolor[rgb]{0.00,0.00,0.81}{#1}}

\newcommand{\KeywordTok}[1]{\textcolor[rgb]{0.13,0.29,0.53}{\textbf{#1}}}
\newcommand{\NormalTok}[1]{#1}
\newcommand{\OperatorTok}[1]{\textcolor[rgb]{0.81,0.36,0.00}{\textbf{#1}}}
\newcommand{\OtherTok}[1]{\textcolor[rgb]{0.56,0.35,0.01}{#1}}

\newcommand{\StringTok}[1]{\textcolor[rgb]{0.31,0.60,0.02}{#1}}

\usepackage{graphicx,grffile}
\makeatletter
\def\maxwidth{\ifdim\Gin@nat@width>\linewidth\linewidth\else\Gin@nat@width\fi}
\def\maxheight{\ifdim\Gin@nat@height>\textheight\textheight\else\Gin@nat@height\fi}
\makeatother
\setkeys{Gin}{width=\maxwidth,height=\maxheight,keepaspectratio}
\makeatletter
\def\fps@figure{htbp}
\makeatother
\usepackage{setspace}
\usepackage{xcolor}
\title{Semiparametric Regression for Dual Population Mortality}
\author{Gary Venter$^1$ and Şule Şahin$^2$\\
 {\small $^1$ Columbia University, Department of Actuarial Sciences, US} \\
   {\small $^2$ University of Liverpool, Department of Mathematical Sciences,\vspace{-0.2cm} }\\
    {\small Institute for Financial and Actuarial Mathematics, UK \vspace{-0.2cm}}\\
   {\small $^2$ Hacettepe University, Department of Actuarial Sciences, Turkey}}
\date{}

\begin{document}
\maketitle

\textbf{Abstract:}\\
Parameter shrinkage applied optimally can always reduce error and
projection variances from those of maximum likelihood estimation. Many
variables that actuaries use are on numerical scales, like age or year,
which require parameters at each point. Rather than shrinking these
towards zero, nearby parameters are better shrunk towards each other.
Semiparametric regression is a statistical discipline for building
curves across parameter classes using shrinkage methodology. It is
similar to but more parsimonious than cubic splines. We introduce it in
the context of Bayesian shrinkage and apply it to joint mortality
modeling for related populations. Bayesian shrinkage of slope changes of
linear splines is an approach to semiparametric modeling that evolved in
the actuarial literature. It has some theoretical and practical
advantages, like closed-form curves, direct and transparent
determination of degree of shrinkage and of placing knots for the
splines, and quantifying goodness of fit. It is also relatively easy to
apply to the many nonlinear models that arise in actuarial work.
\textcolor{black}{We find that it compares well to a more complex state-of-the-art statistical spline shrinkage approach on a popular example from that literature.}

\textbf{Keywords:} Semiparametric regression, Joint mortality, Parameter
shrinkage, Bayesian shrinkage, MCMC

\newpage

\textbf{1 Introduction}\\
Actuaries often need to estimate mortality models for specialized
subpopulations of a larger population, e.g., for pricing products for
high-net-worth individuals. Dual mortality modeling seeks to incorporate
the data of the larger population while allowing the subpopulation model
to differ to the extent it has reliability. We develop a method to apply
Bayesian shrinkage at this step. This lets the parameters of the two
models differ to the extent that this improves the joint posterior
distribution.

The method explored is to use Bayesian semiparametric models for both
populations but make the model for the smaller population the
larger-population model plus a semiparametric model for the differences,
with both fit simultaneously. As an illustration, we apply this to
jointly model male mortality for Denmark and Sweden. This is not a
typical dual-mortality modeling application, as Denmark with 5.6 million
people would often be modeled on its own, even though Sweden is almost
twice as populous. However the data is publicly available and it lets usblack
detail the steps of the process. We build up increasingly complex
models, culminating in a negative binomial version of the
Renshaw-Haberman model (Renshaw and Haberman (2006)).
\textcolor{black}{This adds cohort effects to the }Lee and Carter
(1992)
\textcolor{black}{ model, which lets the time trend vary by age. A generalization by }Hunt
and Blake
(2014)\textcolor{black}{, which has multiple trends each with its own age variation, was tested but did not improve the fit.}

Quite a few actuarial papers have addressed joint age-period-cohort
(APC) modeling of related datasets in reserving and mortality. Li and
Lee (2005) introduce the concept of using a joint stochastic age-period
process for mortality modeling of joint populations, with mean reverting
variations from the joint process for each population. A similar
approach by Jarner and Kryger (2009) models a large and a small
population with the small population having a multi-factor mean-zero
mean reverting spread in log mortality rates. Dowd et al. (2011) model
two populations that can be of comparable or different sizes with mean
reverting stochastic spreads for both period and cohort trends. A
period-trend spread in an APC model is estimated by Cairns et al.
(2011), who use Bayesian Markov-Chain Monte Carlo (MCMC) estimation.
They assume fairly wide priors, not shrinkage priors. This uses the
capability of MCMC to estimate difficult models. Antonio, Bardoutsos,
and Ouburg (2015) estimate the model of Li and Lee (2005) by MCMC.

For some time actuaries have used precursors to semiparametric
regression, like linear and cubic spline curves fit across parameter
types. Semiparametric regression uses parameter shrinkage methodology to
produce an optimized degree of parsimony in the curve construction. The
standard building block for this now is O'Sullivan penalized cubic
splines (O'Sullivan (1986)). Harezlak, Ruppert, and Wand (2018) provides
good background, with R code. Here we use penalized linear splines,
which have some advantages for the models we study.

The actuarial use of shrinkage for linear-spline models started with
Barnett and Zehnwirth (2000), who applied it to APC models for loss
reserving, using an ad hoc shrinkage methodology. Venter, Gutkovich, and
Gao (2019) used frequentist random-effects methods for linear-spline
shrinkage for reserving and mortality models, including joint modeling
of related loss triangles. Venter and Şahin (2018) fit the Hunt-Blake
mortality model with Bayesian shrinkage of linear splines, determining
the shrinkage level by cross-validation. Gao and Meng (2018) used
Bayesian shrinkage on cubic splines for loss reserving models, getting
the shrinkage level by a fully-Bayesian method. Here we use linear
splines on a Renshaw-Haberman model with a fully-Bayesian estimation
approach that combines lasso with MCMC to speed convergence.

Shrinkage methods are discussed in Section 2 and Section 3 has the
application to actuarial models. Section 4 works through the steps for
fitting the joint model, and Section 5 concludes.

\textbf{2 Shrinkage Methodologies}\\
The use of shrinkage-type estimation to improve fitting and projection
accuracy traces back to actuarial credibility theory, then to various
coefficient-shrinkage methods for regression, which we work through in
this section
\textcolor{black}{ to show the reasoning that led to the current model}.
The \textcolor{black}{ evolving methods} have \textcolor{black}{become}
easier to apply, and also have been extended to building curves across
levels of some variables.

\textbf{\emph{2a Credibility}}\\
Shrinking estimates towards the grand mean started with combined
estimation of a number of separate means. A typical example is batting
averages for a group of baseball players. Stein (1956) showed that
properly shrinking such means towards the overall mean always reduces
estimation and prediction variances from those of maximum likelihood
estimation (MLE) when there are three or more means being estimated.
Similar methods have been part of actuarial credibility theory since
Mowbray (1914), and in particular Bühlmann (1967). According to the
Gauss-Markov Theorem, MLE is the minimum-variance unbiased estimate.
Credibility biases the individual estimates towards the grand mean, but
reduces the estimation errors. Thus having more accurate estimates comes
at the cost of possibly asymmetric even though smaller confidence
intervals. This holds in all of the shrinkage methods below.

\textbf{\emph{2b Regularization}}\\
For regression models, Hoerl and Kennard (1970), showed that some degree
of shrinking coefficients towards zero always produces lower fitting and
prediction errors than MLE. This is related to credibility, as in their
ridge-regression approach, variables are typically scaled to have mean
zero, variance one, so reducing coefficients shrinks the fitted means
towards the constant term, which is the overall mean. Their initial
purpose was to handle regression with correlated independent variables,
and they applied a more general procedure for ill-posed problems known
as Tikhonov regularization, from Tikhonov (1943). As a result, shrinkage
methods are also referred to as regularization.

With coefficients \(\beta_j\), ridge regression minimizes the negative
loglikelihood (NLL) plus a selected shrinkage factor \(\lambda\) times
the sum of squares of the coefficients. Thus it minimizes:
\[ NLL+\lambda\sum _j\beta_j^2\]The result of Hoerl and Kennard (1970)
is that there is always some \(\lambda>0\) for which ridge regression
gives a lower estimation variance than does MLE. They did not have a
formula to optimize \(\lambda\), but could find reasonably good values
by cross validation -- that is, by testing prediction on subsamples
omitted from the fitting.

In the 1990s, lasso (see Santosa and Symes (1986) and Tibshirani
(1996)), which minimizes: \[NLL+\lambda\sum _j|\beta_j|\]became a
popular alternative. It shrinks some coefficients exactly to zero, so it
provides variable selection as well as shrinkage estimation. A modeler
can start off with a long list of variables and use lasso to find
combinations that work well together with parameter shrinkage.

\textbf{\emph{2c Random Effects}}\\
Random effects modeling provides a more general way to produce such
shrinkage. Historically it postulated mean-zero normal distributions for
the \(\beta_j\), with variance parameters to be estimated, and optimized
the joint likelihood, which is the product of those normal densities
with the likelihood function. This also pushes the coefficients towards
zero since they start out having mean zero, mode zero distributions.
This can have a different variance for each \(\beta_j\), which can also
be correlated. Ridge regression arises from the special case where there
is a single variance assumed for all the coefficients. Using the normal
distribution for this shrinkage is common but it is not a requirement.
If the double exponential (Laplace) distribution is assumed instead,
lasso becomes a special case as well.

We start by assuming the random effects have independent mean-zero
normal distributions with a common variance. Given a vector \(b\) of
random effects and a vector \(\beta\) of parameters (fixed effects),
with respective design matrices \(Z\) and \(X\), and a vector of
observations \(y\), the model is: \[y = X\beta +Zb +\epsilon\]
\[b \sim N(0,\theta\sigma^2)\] \[\epsilon \sim N(0,\sigma^2)\]

Sometimes fitting the model is described as estimating the parameters
and projecting the random effects. A popular way to fit the model is to
maximize the joint likelihood, which is:
\[p(b,y;\beta)=p(y|\beta,b)p(b)\]This is the likelihood times the
probability of the random effects. The mean-zero distribution for the
\(b\)s pushes their projections towards zero, so this is a form of
shrinkage. Matrix estimation formulas have been worked out for this
similar to those for regression.

Now we consider instead the double exponential, or Laplace, distribution
for the random effects. This is an exponential distribution for positive
values of the variable, and is exponential in \(-b\) for negative
values. Its density is \[p(b|\lambda) = 0.5\lambda e^{-\lambda |b|}\]
Suppose all the random effects have the same \(\lambda\). Then the
negative joint loglikelihood is:
\[-log[p(b,y;\beta)] = log(2) -log(\lambda)+\sum _j \lambda|b_j|+NLL\]The
constants do not affect the minimization, so for a fixed given value of
\(\lambda\), this is the lasso estimation. If \(\lambda\) is considered
a parameter to estimate, then the \(log(\lambda)\) term would be
included in the minimization, giving an estimate of \(\lambda\) as well.
This same calculation starting with a normal prior gives ridge
regression. Thus these are both special cases of random effects, and
both provide estimates of the shrinkage parameter.

Lasso and ridge regression have problems with measures of penalized NLL
like AIC and BIC. These rely on parameter counts, but shrunk parameters
do not use as many degrees of freedom, as they have less ability to
influence fitted values. Because of this, cross validation is popular
for model selection for frequentist shrinkage. There is usually some
degree of subjectivity involved in the choice of cross-validation
approach. We will see below that Bayesian shrinkage has a fast automated
cross-validation method for penalizing the NLL. It provides a good
estimate of what the likelihood would be from a new sample from the same
population, which is the goal of AIC as well.

\textbf{\emph{2d Bayesian Shrinkage}}\\
Bayesian shrinkage is quite similar to random effects, with normal or
Laplace priors used for the parameters like they are used for postulated
distributions of the effects. Ridge regression and lasso estimates are
produced as the posterior modes. The Bayesian and frequentist approaches
are closer than in the past, as random effects are similar to prior
distributions and the Bayesian prior distributions are not necessarily
related to prior beliefs. Rather they are some of the postulated
distributions assumed in the model, just like random effects are, and
they can be rejected or modified based on the posteriors that are
produced.

In the Bayesian approach, a parameter might be assumed to be normally
distributed with mean zero, or Laplace or otherwise distributed. For a
sample \(X\) and parameter vector \(\beta\), with prior \(p(\beta)\) and
likelihood \(p(X|\beta)\), the posterior is given by Bayes Theorem as:
\[p(\beta|X)= p(X|\beta)p(\beta)/p(X)\]Here the probability of the data,
\(p(X)\), is not usually known but it is a constant. Since
\(p(\beta|X)\) must integrate to 1, the posterior is determined by
\(p(X|\beta)p(\beta)\).

Note that if \(\beta\) were random effects, the numerator of
\(p(\beta|X)\) would be the joint likelihood. Thus maximizing the joint
likelihood gives the same estimate as the mode of the posterior
distribution. The main difference between Bayesians and frequentists now
is that the former mostly use the posterior mean, and the latter use the
posterior mode. Bayesians tend to suspect that the mode could be overly
responsive to peculiarities of the sample. Frequentists come from a
lifetime of optimizing goodness of fit of some sort, which the mode
does.

Bayesians generate a sample of the posterior numerically using MCMC.
This ranks parameter samples using the joint likelihood, since this is
proportional to the posterior. MCMC has a variety of sampling methods it
can use. The original one was the Metropolis algorithm, which has a
generator for the possible next sample using the latest accepted sample.
If the new sample parameters have a higher joint probability, it is
accepted. If not, a random draw is used to determine whether it is
accepted or not. An initial group of samples is considered to be warmup
and is eventually eliminated, and the remainder using this procedure
have been shown to follow the posterior distribution.

We feel that random effects estimation can also make use of MCMC to
calculate the posterior mean, but under a different framework. Now we
will just consider the case where there are no fixed effects to estimate
-- all the effects in the model are random effects. A distribution is
postulated for the effects, including a postulated distribution for
\(\lambda\). We know how to calculate the mode of the conditional
distribution of the effects given the data, namely by maximizing the
joint likelihood. By the definition of conditional distributions, the
joint likelihood =
\[p(X|\beta)p(\beta) = p(X, \beta) = p(X)p(\beta|X)\]Thus the
conditional distribution of the effects given the data is proportional
to the joint likelihood, and so can be estimated by MCMC.

What the Bayesians would call the posterior mean and mode of the
parameters can instead be viewed as the conditional mean and mode of the
effects given the data. With MCMC, the effects can be projected by the
conditional mean instead of the conditional mode that was produced by
maximizing the joint likelihood. This does not require Bayes Theorem,
subjective prior and posterior probabilities, or distributions of
parameters, so removes many of the objections frequentists might have to
using MCMC. Admittedly Bayes Theorem is obtained by simply dividing the
definition of conditional distribution by \(p(X)\), but you do not have
to do that division to show that the conditional distribution of the
effects given the data is proportional to the joint likelihood.

\textbf{\emph{2e Goodness of fit from MCMC}}\\
One advantage of using MCMC, whether the mean or mode is used, is that
there is a convenient goodness-of-fit measure. The leave-one-out, or
loo, cross validation method is to refit the model once for every sample
point, leaving out that point in the estimation. Then the likelihood of
that point is computed from the parameters fit without it. The
cross-validation measure, the sum of the loglikelihoods of the omitted
points, has been shown to be a good estimate of the loglikelihood for a
completely new sample from the same population, so it eliminates the
sample bias in the NLL calculation. This is what AIC, etc. aim to do as
well.

But doing all those refits can be very resource intensive. Gelfand
(1996) developed an approximation for an omitted point's likelihood from
a single MCMC sample generated from all the points. He used the
numerical integration technique of importance sampling to estimate a
left-out point's likelihood by the weighted average likelihood across
all the samples, with the weight for a sample proportional to the
reciprocal of the point's likelihood under that sample. That gives
greater weight to the samples that fit that point poorly, and turns out
to be a good estimate of the likelihood that it would have if it had
been left out of the estimation. This estimate for the likelihood of the
point turns out to be the harmonic mean over the samples of the point's
probability in each sample. With this, the MCMC sample of the posterior
distribution is enough to calculate the loo goodness-of-fit measure.

This gives good but volatile estimates of the loo loglikelihood.
Vehtari, Gelman, and Gabry (2017) addressed that by a method similar to
extreme value theory -- they fit a Pareto to the probability reciprocals
and use the Pareto values instead of the actuals for the largest 20\% of
the reciprocals. They call this ``Pareto-smoothed importance sampling.''
It has been extensively tested and is becoming widely adopted. Their
penalized likelihood measure is labeled \(\widehat{elpd}_{loo}\),
standing for ``expected log pointwise predictive density.'' Here we call
\(-\widehat{elpd}_{loo}\) simply loo. It is the NLL plus a penalty, so
is a penalized likelihood
measure\textcolor{black}{, so lower is better}.

The fact that this is a good estimate of the NLL without sample bias
comes with a caveat. The derivation of that assumes that the sample
comes from the modeled process. That is a standard assumption but in
financial areas, models are often viewed as approximations of more
complex processes. Thus a new sample might not come from the process as
modeled. Practitioners sometimes respond to this by using slightly
under-fit models -- that is more parsimonious models with a bit worse
fit than the measure finds optimal.

Using cross validation to determine the degree of shrinkage usually
requires doing the estimation for each left-out subsample for trial
values of the shrinkage constant \(\lambda\). With Bayesian shrinkage
you still have to compute loo for various \(\lambda\)s, but not by
omitting points, as loo already takes care of that. This was the
approach taken by Venter and Şahin (2018).

Another way to determine the degree of shrinkage to use is to put a
wide-enough prior distribution on \(\lambda\) itself. Then the posterior
distribution will include estimation of \(\lambda\). Our experience has
been that the resulting loo from this ends up close to the lowest loo
for any \(\lambda\), and can be even slightly lower, possibly due to
having an entire posterior distribution of \(\lambda\)s. Gao and Meng
(2018) do this, and it is done in the mortality example as well.

Putting a prior on \(\lambda\) is considered to be the fully Bayesian
approach. Estimating \(\lambda\) by cross-validation is not a Bayesian
step. Bayesians often think of a parameter search to optimize a
cross-validation method as a risky approach. A cross-validation measure
like loo is an estimate of how the model would fit a new, independent
sample from the same population, but such estimates themselves have
estimation errors. A search over parameters to optimize the measure thus
runs the risk of producing a measure that is
\textcolor{black}{only the optimum because of the error in estimating the sample bias}.
The fully Bayesian approach avoids this problem. As we saw above, the
fully Bayesian method is also fully frequentist, as it can be done
entirely by random effects with a postulated distribution for
\(\lambda\).

\textbf{\emph{2f Postulated shrinkage distributions}}\\
The shape of the shrinkage distribution can also have an effect on the
model fit. The Laplace distribution is more peaked at zero than the
normal, and also has heavier tails, but is lighter at some intermediate
values. An increasingly popular shrinkage distribution is the
Cauchy\[1/p(\beta) = \pi (1+\lambda^2\beta^2)/\lambda\]
\[-log(p(\beta)) = -log\lambda + log \pi +log(1+\lambda^2\beta^2)\]It is
even more concentrated near zero and has still heavier tails than the
Laplace. For a fixed \(\lambda\), the constants in the density drop out,
so the conditional mode minimizes:
\[NLL+\sum log(1+\lambda^2\beta_j^2)\]This is an alternative to both
lasso and ridge regression. Cauchy shrinkage often produces more
parsimonious models than the normal or Laplace do. It can have a bit
better or bit worse penalized likelihood, but even if slightly worse,
the greater parsimony makes it worth considering. It also seems to
produce tighter ranges of parameters.

The Cauchy is the t-distribution with \(\nu = 1\). If \(\nu = 2\), the
conditional mode minimizes \[NLL+1.5\sum log(2+\lambda^2\beta_j^2)\]If
\(\nu = 3\), this becomes: \[NLL+2\sum log(3+\lambda^2\beta_j^2)\]The
normal is the limiting case of t distributions with ever larger degrees
of freedom. Degrees of freedom \(\nu\) does not have to be an integer,
so a continuous distribution postulated for \(\nu\) could be used to get
an indication for how heavy-tailed the shrinkage distribution should be
for a given data set. Actually, the Laplace distribution can be
approximated by a t distribution. A t with \(\nu = 6\) matches a Laplace
with scale parameter \(\sqrt{3}/2\) in both variance and kurtosis (and
since the odd moments are zero, matches all 5 moments that exist for
this t), so is a reasonable approximation. \(E(1/X, X \ne 0)\) is zero
for this t but is undefined for the Laplace, which is due to its sharp
point at zero.

We test priors for the example in Section 4. The Laplace is assumed
while building the model, but afterwards the shrinkage distribution is
tested by using a students-t shrinkage distribution with an assumed
prior for \(\nu\), and finding the conditional distribution of \(\nu\)
given the data. For this case, the conditional mean is close to
\(\nu = 2\). This is heavier-tailed than the Laplace but less so than
the Cauchy. The t-2 distribution, with \(\nu = 2\), is a closed-form
distribution: \[f(\beta)= (2+\beta^2)^{-1.5}\]
\[F(\beta)= [\beta /\sqrt{2+\beta^2}+1]/2\]Still the Cauchy is a viable
alternative to get to a more parsimonious model.

The versatility of the t for shrinkage also opens up the possibility of
using different but correlated shrinkage for the various parameters, as
the multi-variate t prior is an option in the software package we use.
For \(\nu = 2\), the covariances are infinite, but a correlation matrix
can still be input. Two adjacent slope changes are probably quite
negatively correlated, as making one high and one low can often be
reversed with little effect on the outcome. Recognizing this correlation
in the assumed shrinkage distribution might improve convergence. We
leave this as a possibility for future research.

\textbf{\emph{2g Semiparametric regression}}\\
Harezlak, Ruppert, and Wand (2018) represents the contemporary view of
semiparametric methodology. It uses a few advanced methods but is not
easy to adapt to the fully Bayesian approach to shrinkage. We compare
our simpler method of Bayesian shrinkage on the slope changes of linear
splines to what they are doing and find the methods comparable in
results, with our method having the advantages of a Bayesian approach,
such as being able to use the mean of the conditional distribution of
\(\lambda\) given the data instead of using the chancy cross validation
method. They provide an oft-cited example of using semiparametric
methods to create a curve for mean by year of construction for the floor
area a fixed monetary unit could get for apartments in Warsaw in 2007 --
2009. We fit their data by using Bayesian shrinkage on the slope changes
of linear splines (piecewise linear curves). Figure 1 shows the data and
the two resulting curves, which are quite similar.

Historically the spline-fitting literature rejected linear splines on
aesthetic grounds for being too jagged. But this was before shrinkage.
Shrinkage can be done on cubic splines by penalizing the average second
derivative of the spline curve (see Hastie, Tibshirani, and Friedman
(2017)), much like ridge regression penalizes the sum of squared
parameters. The actuarial approach to linear-spline shrinkage, following
Barnett and Zehnwirth (2000), analogously shrinks the slope changes,
which are the second differences of the splines. As Figure 1 shows, this
does not differ much in jaggedness from the cubic-spline fit. Another
advantage is that the knots of the linear splines come out directly from
the estimation as the points with zero slope change and so are
determined simultaneously with the fits. This is a separate procedure
with cubic splines, which makes the overall optimization more awkward.

Harezlak, Ruppert, and Wand (2018) recommend O'Sullivan penalized
splines. These are a form of smoothing splines that meet a roughness
penalty controlled by a smoothing constant \(\lambda\). They pick this
\(\lambda\) using ``generalized cross validation'' from Craven and Wahba
(1979). These splines are averages of a number of cubic functions and
are not closed form, but there is available software to calculate them.
They also discuss how to pick the number of knots \(K\) for tying
together the splines. They feel that with the smoothing done anyway, 35
knots spread out by the quantiles of the data will usually suffice. They
do have some functions for testing the number of knots, however.

Actuarial smoothed linear splines can be done on any variables that can
be ordered numerically 1, 2, 3, \ldots. This is natural for years, ages,
etc. but can make sense for class variables like vehicle use classes
that have some degree of ranking involved. The parameter shrinkage is
applied to the slope changes between two adjacent linear segments.
Initially slope changes are allowed between all adjacent variables, but
if the shrinkage reduces a slope change to zero, the line segment simply
continues at that point, and a knot is eliminated.

Thus the shrinkage chooses the knots, which can then determine how many
are needed and put them where the curves most need to change. This is an
advantage over the O'Sullivan splines, which usually use a fixed number
like 35 knots spread according to the number of observations, not
changes in the levels. In our experience, 35 knots is usually enough,
but can sometimes lead to over-parameterization. The degree of shrinkage
could be selected by loo cross validation, which seems as good or better
than generalized cross validation once you have a posterior distribution
available, but we use the fully Bayesian method of letting the
conditional distribution of the degree of shrinkage given the data
determine this. Bayesians tend to prefer this approach, as discussed at
the end of 2e above. The key to making this work is that you can create
a design matrix with slope-change dummy
variables\textcolor{black}{, as discussed in Section 3 below}.

A useful short-cut to the fitting when there are a lot of variables is
to use lasso in an initial sorting out process. The R lasso program
glmnet \textcolor{black}{developed by} Friedman, Hastie, and Tibshirani
(2010) has a built-in cross-validation function
cv.glmnet()\textcolor{black}{, which we use with all its standard defaults}.
It uses its own cross-validation method to select a range of \(\lambda\)
values that are worth considering. We take the low end of this range,
which eliminates the fewest variables, as an initial selection point.
Usually more variables are worth taking out, but this can happen later.
Adjacent dummy variables are highly negatively correlated, and one can
usually substitute for another with little change in the overall fit.
This creates difficulties for MCMC, but lasso can quickly sort out the
variables with least overall contribution to the model, and then MCMC
can readily deal with those that remain. We did this in the Warsaw
example and also in the mortality example below. Even if by chance a
slope change is left out that would add value to the final model, the
impact is likely to be minimal.

An advantage of this method is that the fully Bayesian approach chooses
the knots and the smoothing constant as parts of the posterior
distributions, and as noted above, fully Bayesian has less risk of
producing a model with an exaggerated fit measure than cross validation.
Also being able to do this with just a design matrix and MCMC fitting
software allows flexibility in model construction. More complex bilinear
models like Lee-Carter, Renshaw-Haberman, and Hunt-Blake can be readily
fit, as we shall see below. Further, it is straightforward to use any
desired parameter distribution, like negative binomial in the example.

\begin{figure}
\centering
\includegraphics{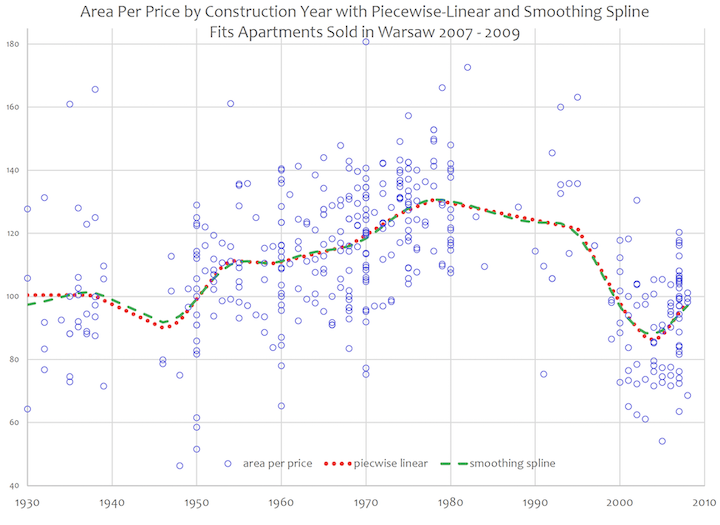}
\caption{Semiparametric Fits to Warsaw Apartment Data \label{figa}}
\end{figure}

\textbf{3 Joint Mortality Modeling by Shrinkage}\\
We are using data from the Human Mortality Database (2019), which
organizes tables by year of death and age at death. The year of birth is
approximately year of death minus age at death, depending on whether the
death is before or after the birthday in the year. For convenience, we
consider cohort to be year of death minus age at death. Cohorts are
denoted as \(n\), with parameters \(p[n]\), age at death by \(u\), with
parameters \(q[u]\), so the year of death is \(w=n+u\), with parameters
\(r[n+u]\).
\textcolor{black}{The data we use is, as is common, in complete age by year rectangles with $a$ years and $b$ ages, and $ab$ cells. These points are from $a+b-1$ cohorts, which have from 1 to $min(a,b)$ ages in them. Any two of the indices $n,u,w$ will identify a cell. We use $n,u$, which avoids subtraction of indices.}
Also \(c\) is a constant term which is not shrunk in the modeling. The
parameters \textcolor{black}{add and multiply} in the models to give
the log of the mortality rate.

Initially the number of deaths is assumed Poisson in the mortality rate
times the exposures, which are the number living at the beginning of the
year. Theoretically, deaths counts are the sum of Bernoulli processes,
and so should be binomial, but for small probabilities, the binomial is
very close to the Poisson, which is sometimes easier to model. Later we
test the Poisson assumption against heavier-tailed distributions.

\textbf{\emph{3a Mortality models}}\\
A basic model is the linear APC
model.
This has become fairly popular with actuaries and it is our starting
point, as it is a purely linear model that can be estimated with
standard packages, including lasso packages. Its
\textcolor{black}{log mortality rate $m[n,u]$, for cohort $n$ and age $u$}
is given by: \[m[n,u] = c+p[n]+q[u]+r[n+u]\] With exposures e{[}n,u{]},
the deaths \(y[n,u]\) are
\textcolor{black}{often assumed to be Poisson distributed:$$y[n,u] \sim Poisson\left(e[n,u]e^{m[n,u]}\right)$$}The
model of Renshaw and Haberman (2006) adds cohorts to the Lee and Carter
(1992) model, which itself allows the time trend \(r[n+u]\) to vary by
age, reflecting the fact that medical advances, etc. tend to improve the
mortality rates for some ages more than others. Renshaw and Haberman
(2006)
\textcolor{black}{ also include trend weights on the cohort effects.}
We have tried that for other populations and found that the age effects
were not consistent from one cohort to another. This aspect does not
appear to have been widely used. Haberman and Renshaw (2011) do find
that it improves the loglikelihood somewhat in one of three datasets
they look at. Chang and Sherris (2018) graph the age effect for several
cohorts of their data, which shows some variations among cohorts. They
find that a two-parameter model for age variation within each cohort
captures this well. We did not try this model. A more generalized
underlying model with this same adjustment is in Xu, Sherris, and Ziveyi
(2019).

Denoting the trend weight for age \(u\) by \(s[u]\), the
Renshaw-Haberman model we use here is:
\[m[n,u] = c+p[n]+q[u]+s[u]r[n+u]\] Hunt and Blake (2014) allow the
possibility of different trends with different age weights either
simultaneously or in succession. This addresses a potential problem with
the Lee-Carter model: the ages with the most mortality improvement can
change over time. Venter and Şahin (2018) suggest an informal test for
the need for multiple trends, and find that they indeed help in modeling
the US male population for ages 30--89. In the example here we fit the
model to ages 50 and up, and did this test, which suggested that one
trend is sufficient, so the model is presented in that form. It is then
the same as the special case of the Renshaw-Haberman model with no
age-cohort interaction. In other populations we have noticed that
modeling a wide range of ages over a long period can benefit from
multiple trends.

The Renshaw-Haberman and Lee-Carter models are sometimes described as
bilinear APC models, as they are linear if either the \(s\) or \(r\)
parameters are taken as fixed constants. The age weights help model the
real phenomenon of differential trends by age, but their biggest
weakness is that they are fixed over the observation period, and the
ages with the most trend can change over time as medical and public
health trends change. The possibility of multiple trends in the
Hunt-Blake model can reflect such changes.

\textbf{\emph{3b Design matrix}}\\
To set the basic APC model up to use linear-modeling software packages,
the entire array of death counts \(y[n,u]\) is typically strung out into
a single column \(d[j]\) of length \(ab\).
It is generally helpful to make parallel columns to record which age, period, and
cohort cell \(d[j]\) comes from. Then a design matrix can be created
where all the variables \(p[n],q[u],r[n+u]\) have columns parallel to
the death counts, and are represented as 0,1 dummy variables that have a
1 only for the observations that variable applies to.
In practice we keep the constant \(c\) out of the design matrix, and also
leave out the first cohort, age and year variables \(p[1],q[1],r[1]\).

Let \(\beta\) be the vector of parameters and \(\widetilde{e}\) be the
ab-vector of exposures corresponding to \(d\). Then the ab-vector
\(\widehat{m}\) of estimated \textcolor{black}{log} mortality rates is
\[\hspace{13em}\widehat{m} = c+X\beta \hspace{13em} (i)\]and
\[\hspace{10em}d[j] \sim Poisson\left(\widetilde {e}[j]e^{\widehat{m}[j]}\right)\hspace{10em} (ii)\]Both
formulas \((i)\: ,\: (ii)\) still hold when \(X\) is singular,
\textcolor{black}{which can happen because of APC redundancy prior to shrinkage.}
In that case you cannot estimate \(\beta\) as \((X'X)^{-1}X'd\), but
however it is estimated, the fitted values are computed using
\((i)\: ,\: (ii)\). These are the sampling statements that can be given
to MCMC to create samples of the conditional distribution of the
parameters given the data.

In the semiparametric case, where the parameters are slope changes to
linear splines across the cohort, age, and year parameters a different design matrix
is needed.
\textcolor{black}{ Slope changes are second differences in the piecewise-linear curves fit across each parameter type (age, period, etc.). The slope-change variable for age $i$, for example, is zero at all points in $d$ that are for ages less than $i$. It is 1 for all points for age $i$. The first differences have been increased by that parameter at that point as well. So any cell for age $i+1$ gets the first difference plus the level value at age $i$, so it gets two summands of the age $i$ variable. Thus the dummy for that variable is 2 at those points. Then it is 3 at points for age $i+2$, etc. In general, the age $i$ dummy variable gets value $(k-i+1)_+$ at points for age $k$. The same formula works for the year $i$ variable at points from year $k$, etc.}

In the original levels design matrix, each row has three columns with a
1 in them. When a row is multiplied by \(\beta\), those three values
would pick out the year, age, and cohort parameters for that cell, and
they would be added up to get the \(\widehat{m}\) value for the cell. In
principle this is the same for the slope-change design matrix, but for
any given row there would be non-zero elements for every year, age, and
cohort up to that point. Those values would multiply the slope-change
parameters in \(\beta\) by the number of times that they add up for that
cell, and those would be summed to give \(\widehat{m}\).

\textbf{\emph{3c Fitting procedure}}\\
For the two population case, assume that the dependent variable \(d\)
column has all the observations for the first population followed by all
the observations for the second population
\textcolor{black}{in the same order}. Our approach is to fit a
semiparametric model to the larger population and simultaneously fit a
semiparametric model for the differences in the second population from
the first. Then the smaller population parameters (in log form) are the
sum of the two semiparametric models. With shrinkage applied, the
difference parameters hopefully are smaller, so the second population
would use fewer effective parameters than the first.
\textcolor{black}{The design matrix for the semiparametric fit to the second-population differences is the same as the design matrix for either of the population's log mortality rates. They all specify how many times each slope change parameter will be added up for each fitted value. The parameter vector will have parameters for the first population's slope changes, followed by parameters for the slope change differences for the second population from the first. Because the constant term is not shrunk, it is not in the design matrix but is added in separately. The second-population design matrix does include a constant variable for the difference, and that is shrunk. (In the symbolic equation below, the design matrix includes both constants.)}

%
%
%

\textcolor{black}{The Poisson mean vector is the element-by-element multiplication of the exposure vector with the exponentiation of the matrix multiplication of the design matrix by the parameter vector. This equation provides a visualization of the APC model with $k$ parameters:} 
\begin{eqnarray*} 
\label{eq:apc1}
{\large E} \tiny{\overbrace{\left(
\begin{array}{l}
\\
\text{First}\\
 \text{Pop.}\\
\text{Counts}\\
 \\\\\\\hdashline[3pt/3pt]
\\\\
\text{Second} \\
\text{Pop.}\\
\text{Counts} \\\\
\\
\end{array}
\right)}^{\text{Death counts}\ (\tilde{d})}}_{\hspace{-0.09in} 2ab\times 1} \hspace{-0.17in} {\large =} \hspace{0.2in}
\tiny{\overbrace{\left(
\begin{array}{l}
\\
\text{First} \\
\text{Pop.}\\
\text{Expo-} \\
\text{sures} \\\\\\\hdashline[3pt/3pt]
\\\\
\text{Second} \\
\text{Pop.}\\
\text{Expo-} \\
\text{sures} \\
\\
\end{array}
\right)}^{\text{Exposures} \ (\tilde{e})}}_{2ab\times 1}
\hspace{-0.2in}\circ \hspace{0.1in}{\normalsize exp} \hspace{0.2in} \tiny{\overbrace{\left(
\begin{array}{c|c}
\\
\text{All APC Variables} & \text{All Zeros for} \\
\text{Slope Change} &\text{First} \\
\text{Design Matrix for}& \text{Population}\\
\text{First}& \\
\text{Population}\\\\\hdashline[3pt/3pt]
\\
\text{Same Matrix to Give}&\text{Same Design Matrix}\\
\text{First Pop.}&\text{But Now for} \\
\text{Model to}&  \text{Difference Parameters}\\
\text{Second}&\text{for Slope Changes}\\
\text{Population}& \text{for the Second Pop.}\\
 \\
\end{array}
\right)}^{\text{Design matrix}}}_{2ab\times k} \hspace{-0.2in}\times 
\tiny{\overbrace{\left(
\begin{array}{l}
\\
\text{First} \\
\text{Pop.} \\
\text{Param.}\\\\\hdashline[3pt/3pt]
\\
\text{Second} \\
\text{Pop.} \\
\text{Param.}
\\\\
\end{array}
\right)}^{\text{Parameter matrix}}}_{\hspace{-0.09in}k\times 1}
\end{eqnarray*}

\normalsize

The overall combined design matrix \textcolor{black}{has four quadrants. All the age, period and cohort variables have columns for the first population and are repeated as columns for the second population's difference variables. The first half of the rows are for the first population and the second half is for the second population. The upper left quadrant is just the single population design matrix. The upper right quadrant is all zero, so the first population uses only its own variables. The lower left quadrant repeats the upper left quadrant, so the second population gets all the parameter values from the first population. The same single population design matrix is repeated in the lower right quadrant for the differences. Now this applies to the slope change variables for the second population differences. That population gets the sum of the parameters for the first population plus parameters of its own differences. If the difference parameters were not shrunk, the second population would end up with its own mortality parameters. If those parameters are all zero, the two populations would get the same parameters. The shrinkage lets the second population use the parameters from the first to the extent that they are relevant, and then adds in changes as might be needed.} \textcolor{black}{Although the second population columns start off the same as first, in the course of the fitting some variables with low t-statistics are forced to have parameters of zero by taking their columns out of the design matrix, and these variables could be different in the two populations.}

MCMC is an effective way to search for good parameters, but these
variables make it difficult. Consecutive variables are highly correlated
-- here many are 99.9\% correlated. There could be hundreds of variables
in APC models for reasonably large datasets. It could take days or weeks
to converge to good parameters running MCMC software on all of those
correlated parameters. To speed this up, we start with lasso, which is
usually very fast, to identify variables that can be eliminated, and
then use a reduced variable set for MCMC. Lasso software, like the R
package glmnet, includes cross validation routines to help select
\(\lambda\). This gives a range of \(\lambda\) values that is worth
considering. We use the lowest \(\lambda\) value this suggests, which
leaves the largest number of variables, to generate the starting
variables for MCMC. Usually this still includes more parameters than are
finally used in the model.

When a slope-change variable is taken out, the slope of the fitted
piecewise linear curve does not change at that point, which extends the
previous line segment. If instead a small positive or negative value
were to be estimated, the slope would change very slightly at that
point, probably with not much change to the fitted values. When lasso
cross validation suggests that the slope change is not useful, that is
saying it would not add to the predictive value of the model. It is
possible that a few of the omitted points in the end could have improved
the model, but probably they would not have made a lot of difference. So
even with MCMC, there is no guarantee that all the possible models have
been examined and the very best selected. Nonetheless this procedure is
a reasonable way to get very good models.

Constraints are needed to get convergence of APC models. First of all,
with the constant term included, the first age, period, and cohort
parameters are set to zero. This is not enough to get a unique solution,
however. With age, period, and cohort parameters all included, the
design matrix is singular.
\textcolor{black}{The parameters are not uniquely determined in APC models in general: you can increase the period trend and offset that by decreasing the cohort trend, with adjustments as needed to the age parameters, and get exactly the same fitted values. Additional constraints can prevent this. A typical constraint is to force the cohort parameters to sum to 1.0.}
Venter and Şahin (2018)
\textcolor{black}{constrain the slope of the cohort parameters to be 0, which makes the period trend the only trend so it can be interpreted as the trend given that there is no cohort trend. They also constrain the age weights on trend in Lee-Carter, etc. to be positive with the highest value $=1$. This then makes the trend interpretable as the trend for the age with the highest trend. Another popular constraint is to instead make the trend weights sum to 1.0.}

\textcolor{black}{Some analysis of APC models takes the view that there are no true cohort and period trends since they only become specific after arbitrary constraints. But often there are things known or knowable about the drivers of these trends, like behavioral changes across generations, differences in medical technology and access to health care, etc., and these imply differences between time trend and cohort trend. Also, in Bayesian shrinkage and random effect models, two parameter sets with all the same fitted values will not be equally likely. Their probabilities are proportional to the joint likelihood, which is the likelihood of the data times the postulated (or prior) distribution of the effects (or parameters). Since these are shrinkage distribution, a parameter closer to zero will have higher probability. Over a number of parameters, these will never come out identical across the entire sample. Even though the design matrix is singular, there are still unique parameter sets with the highest probability.}

\textcolor{black}{It has been our experience that letting the parameters sort themselves out with no constraints usually gives a bit better fit, by loo, than you can get with the constraints. Parameter sets with lower prior probabilities show up less frequently in the posterior distribution, so the posterior mean gives a model with parameters closer to zero. Such a model would usually have a better fit by loo penalized likelihood even though having the same loglikelihood as other fits. The loo penalty does not look at the size of parameters -- just at the out-of-sample predictions -- but these are better with properly shrunk parameters. You could say that fitting by shrinkage is itself a complicated type of constraint that does give unique parameters, but that these parameters do not have the ready interpretations that some constraints imply. Still, the better predictive accuracy measured by loo is some evidence, even if not totally definitive, that the trends estimated without other constraints better correspond to empirical reality. In this study we make the age weights on the period trend positive with a maximum of 1.0, but only constrain the trends themselves by Bayesian shrinkage.}

Using slope change variables, with some eliminated, helps with
identifiability. If a straight regression is done for a simple
age-period model, the slope change design matrix gives exactly the same
\(q,r\) parameter values that the 0,1 dummy matrix gives, and the APC
matrix is still singular. But when enough slope changes are set to zero,
the matrix is no longer singular. Our procedure, then, is to use lasso
for an AP design matrix, take out the slope change variables that lasso
says are not needed, then add in all the cohort variables, and run lasso
again. The maximum variable set lasso finds for this is then the
starting point for MCMC, without any other constraints on the variables.

\begin{table}[]
\centering
\caption{Excerpt of Design Matrix}
\begin{tabular}{cccrrrrrrr}
cohort & year & age & deaths & r[2] & r[3] & r[4] & q[2] & q[3] & q[4] \\
1920   & 1970 & 50  & 305    & 0   & 0   & 0   & 37  & 36  & 35  \\
1919   & 1970 & 51  & 306    & 1   & 0   & 0   & 36  & 35  & 34  \\
1918   & 1970 & 52  & 344    & 2   & 1   & 0   & 35  & 34  & 33  \\
1917   & 1970 & 53  & 359    & 3   & 2   & 1   & 34  & 33  & 32  \\
1916   & 1970 & 54  & 384    & 4   & 3   & 2   & 33  & 32  & 31  \\
1915   & 1970 & 55  & 417    & 5   & 4   & 3   & 32  & 31  & 30  \\
1914   & 1970 & 56  & 496    & 6   & 5   & 4   & 31  & 30  & 29  \\
1913   & 1970 & 57  & 570    & 7   & 6   & 5   & 30  & 29  & 28  \\
1912   & 1970 & 58  & 569    & 8   & 7   & 6   & 29  & 28  & 27  \\
1911   & 1970 & 59  & 605    & 9   & 8   & 7   & 28  & 27  & 26  \\
1910   & 1970 & 60  & 732    & 10  & 9   & 8   & 27  & 26  & 25  \\
1909   & 1970 & 61  & 781    & 11  & 10  & 9   & 26  & 25  & 24  \\
1908   & 1970 & 62  & 793    & 12  & 11  & 10  & 25  & 24  & 23  \\
1907   & 1970 & 63  & 854    & 13  & 12  & 11  & 24  & 23  & 22  \\
1906   & 1970 & 64  & 953    & 14  & 13  & 12  & 23  & 22  & 21  \\
1905   & 1970 & 65  & 943    & 15  & 14  & 13  & 22  & 21  & 20  \\
1904   & 1970 & 66  & 1003   & 16  & 15  & 14  & 21  & 20  & 19  \\
1903   & 1970 & 67  & 1092   & 17  & 16  & 15  & 20  & 19  & 18  \\
1902   & 1970 & 68  & 1156   & 18  & 17  & 16  & 19  & 18  & 17  \\
1901   & 1970 & 69  & 1192   & 19  & 18  & 17  & 18  & 17  & 16  \\
1900   & 1970 & 70  & 1266   & 20  & 19  & 18  & 17  & 16  & 15  \\
1899   & 1970 & 71  & 1237   & 21  & 20  & 19  & 16  & 15  & 14  \\
1898   & 1970 & 72  & 1324   & 22  & 21  & 20  & 15  & 14  & 13  \\
1897   & 1970 & 73  & 1338   & 23  & 22  & 21  & 14  & 13  & 12  \\
1896   & 1970 & 74  & 1414   & 24  & 23  & 22  & 13  & 12  & 11  \\
1895   & 1970 & 75  & 1317   & 25  & 24  & 23  & 12  & 11  & 10  \\
1894   & 1970 & 76  & 1481   & 26  & 25  & 24  & 11  & 10  & 9   \\
1893   & 1970 & 77  & 1363   & 27  & 26  & 25  & 10  & 9   & 8   \\
1892   & 1970 & 78  & 1323   & 28  & 27  & 26  & 9   & 8   & 7   \\
1891   & 1970 & 79  & 1311   & 29  & 28  & 27  & 8   & 7   & 6   \\
1890   & 1970 & 80  & 1294   & 30  & 29  & 28  & 7   & 6   & 5   \\
1889   & 1970 & 81  & 1262   & 31  & 30  & 29  & 6   & 5   & 4   \\
1888   & 1970 & 82  & 1187   & 32  & 31  & 30  & 5   & 4   & 3   \\
1887   & 1970 & 83  & 1189   & 33  & 32  & 31  & 4   & 3   & 2   \\
1886   & 1970 & 84  & 1075   & 34  & 33  & 32  & 3   & 2   & 1   \\
1885   & 1970 & 85  & 1024   & 35  & 34  & 33  & 2   & 1   & 0   \\
1884   & 1970 & 86  & 842    & 36  & 35  & 34  & 1   & 0   & 0   \\
1883   & 1970 & 87  & 764    & 37  & 36  & 35  & 0   & 0   & 0   \\
1921   & 1971 & 50  & 334    & 0   & 0   & 0   & 38  & 37  & 36  \\
1920   & 1971 & 51  & 352    & 1   & 0   & 0   & 37  & 36  & 35  \\
1919   & 1971 & 52  & 330    & 2   & 1   & 0   & 36  & 35  & 34  \\
1918   & 1971 & 53  & 363    & 3   & 2   & 1   & 35  & 34  & 33 
\end{tabular}
\end{table}

The MCMC runs usually estimate a fair number of the remaining parameters
as close to zero but with large standard deviations. This is analogous
to having very low t-statistics. We take out those variables as well,
but look at how loo changes. As long as loo does not get worse from
this, those variables are left out. Usually a fair number of the
variables that lasso identified are eliminated in this process. Leaving
them in has little effect on the fit or the loo measure, but the data is
more manageable without them. We use the best APC model produced by this
as the starting point for the bi-linear models, which add age modifiers
to the period trends.

The steps we used to fit an APC model are then:

\begin{enumerate}
  \item Make a slope-change design matrix for the age and period variables and another one for the cohort variables.
  \item Do a lasso run \textcolor{black}{with the age and period variables only} using cv.glmnet in default mode and list the coefficients for lambda.min. Take the variables with coefficients of zero out of the age-period design matrix.
  \item Add the cohort design matrix in with the remaining columns.
  \item Run lasso again with cv.glmnet and list the coefficients for lambda.min. Take the variables with coefficients of zero out of the combined design matrix.
  \item Run MCMC beginning with this design matrix and compute the loo fit measure. \textcolor{black}{This is the end of the formal fitting procedure for the APC model, but next the low-t variables are taken out by the modeler as desired. The rule we used is like taking out the variables that reduce r-bar-squared in a standard regression, but is similarly problematic in using a fit measure to remove the variables. The following outlines how we did it, but other modelers may have their own approaches to this.}
  \item Look at the MCMC output and identify variables with high standard deviations and means near zero.
  \item Take out these variables, run MCMC again, and compute loo.
  \item If loo has improved, see if there are more variables to take out, and repeat. If loo is worse, put some variables back in.
\end{enumerate}

\textbf{4 Swedish-Danish Male Mortality Example}\\
We model deaths from 1970-2016 for ages 50-99 and cohorts 1883-1953 from
the Human Mortality Database. Figure 2 graphs the cumulative mortality
trends by age ranges for both populations to see if there are any
apparent shifts in the ages with the highest trend. The trends are a bit
different in the two countries, but they look a lot like what the
Lee-Carter model assumes, with the relative trends among the ages
reasonably consistent over time.

\begin{figure}
\centering
\includegraphics[width=0.9\textwidth,height=\textheight]{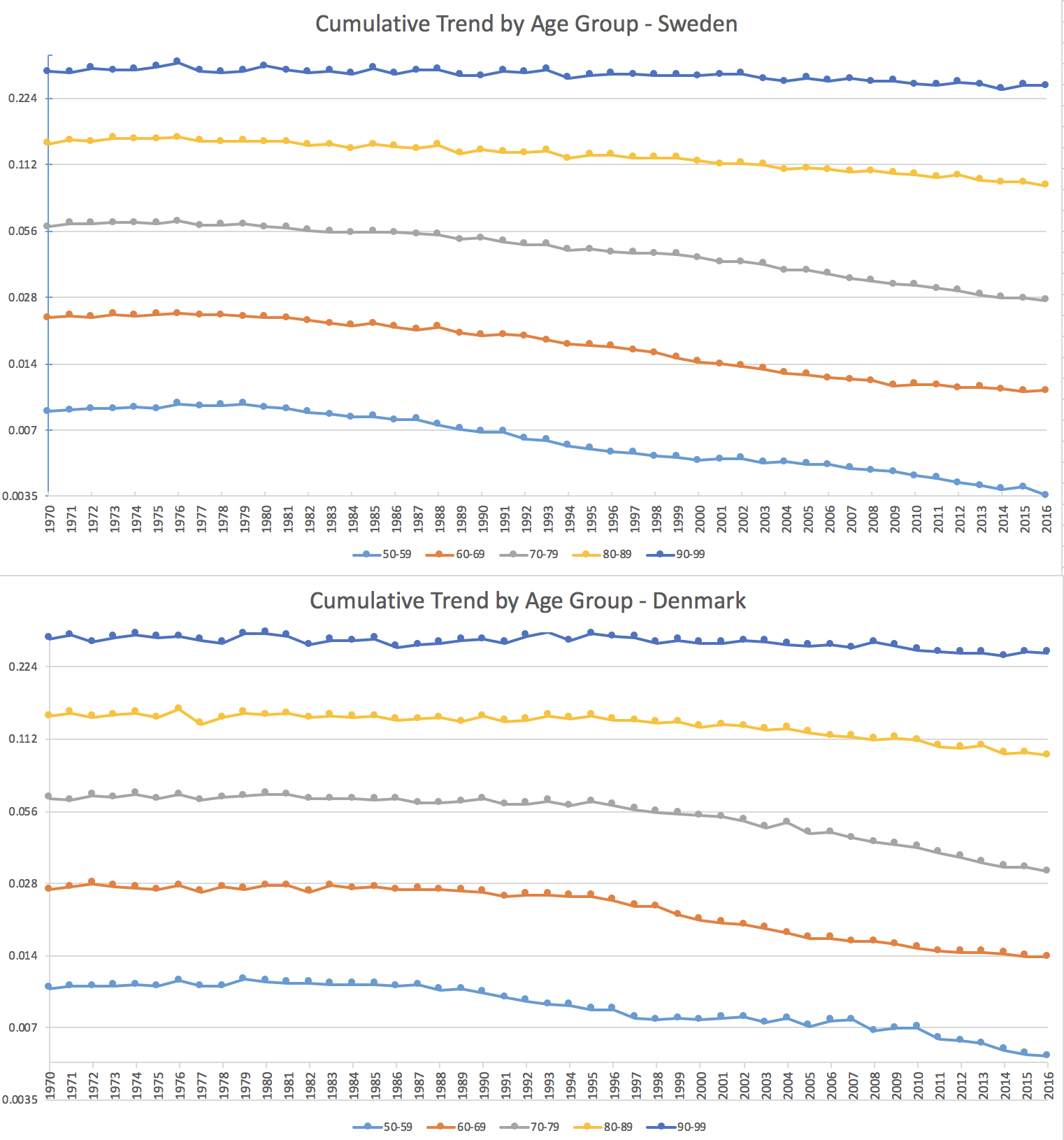}
\caption{Cumulative Trends by Age Group \label{fig1}}
\end{figure}

\textbf{\emph{4a APC model}}\\
Each population has 50 ages for 47 years coming from 71 cohorts. Leaving
out the first age, period, and cohort of each population, and adding the
dummy variable for Denmark, gives 331 variables for the APC design
matrix. Table 1 has an excerpt from the design matrix, showing the first
few age and cohort variables for Sweden for several observations.

We started with a lasso run, using the R GLM lasso package glmnet,
without the cohort variables. The cv.glmnet app does the lasso
estimation as well as this package's default cross validation. One of
the outputs is the smallest \(\lambda\) (thus the one with the most
non-zero coefficients) that meets their cross-validation test. We ran
the package twice. The first time was without the cohort variables, and
then with the resulting non-zero variables from the first run plus all
the cohort variables.

This resulted in 80 variables: 22 age variables, 16 year variables, and
42 cohort variables remained after lasso. We use the Stan MCMC
application, specifically the R version, rstan provided by Stan
Development Team (2020). This requires specifying an assumed
distribution for each parameter, and it simulates the conditional
distribution of the parameters given the data. We used the double
exponential distribution for all the slope-change variables. This is a
fairly common choice, usually called Bayesian lasso, but we revisit it
below. The double exponential distribution in Stan has a scale parameter
\(s = 1/\lambda > 0\). An assumed distribution was provided for that, as
well as for the constant term, \textcolor{black}{as discussed below}.

Eliminating the zero-coefficient variables enabled the APC model to be
fit with no further constraints. Usually something like forcing the
cohort parameters to have zero trend is needed for identifiability. That
particular constraint means that the period trends can be interpreted as
the trends given that they have all the trend and none is in the
cohorts. The method here is simpler and gives a bit better fit, but does
not allow for such an easy interpretation.

For positive parameters, we prefer to start with a distribution
proportional to \(1/x\). This diverges both as \(x\) gets small and as
it gets large, so it gives balancing strong pulls up and down, whereas a
positive uniform distribution diverges upwardly only, and this can bias
the parameter upward. As an example where the integral is known,
consider a distribution for the mean \(\beta\) of a Poisson distribution
with probability function proportional to \(e^{-\beta}\beta^k\). With a
uniform distribution for \(\beta\), which is proportional to 1, the
conditional distribution of \(\beta\) given an observation \(k\) is also
proportional to \(e^{-\beta}\beta^k\), which makes it a gamma in \(k+1\)
and 1, with mean \(k+1\). But if the distribution of \(\beta\) is
assumed proportional to \(1/\beta\), the conditional is proportional to
\(e^{-\beta}\beta^{k-1}\), which makes it a gamma in \(k,1\), with mean
\(k\). Thus the \(1/\beta\) unconditional distribution take the data at
face value, whereas the uniform pushes it upwards. Numerical examples
with other distributions find similar results.

The \(1/\beta\) assumption is easier to implement in Stan by assuming
\(log(\beta)\) is uniform. Stan just assumes a uniform distribution is
proportional to a constant, so can be ignored. If not specified, it
takes the range \(\pm 1.7977\cdot10^{308}\), which is the largest
expressible in double-precision format. We usually start with a fairly
wide range for uniform distributions, but if after working with the
model for a while the conditional distributions tend to end up in some
narrower range, we tend to aim for something a bit wider than that range
-- mostly to save the program from searching among useless values. We
ended up with assuming that the log of the constant is uniform on
{[}-8,-3{]}, with \(log(s)\) uniform on {[}-6,-3{]}.
\textcolor{black}{The parameter distributions ended up well within these ranges. Note that this happened to be set up as the constant for the log of the mortality, and so is exponentiated twice in the final model, but could have just as well have been set up as the log of the final constant.}

The best purely APC model had 54 parameters, a loo measure of 22,158.5,
and NLL of 22,081.7. The sample-bias penalty, 76.8, is the difference
between these. This penalty is a bit high for 54 parameters with
shrinkage, which veteran Stan modelers tend to find suspicious in terms
of predictive ability of the model. We know from the preliminary visual
test that there is some variation in trend across the ages, and not
reflecting this could be the problem.

\begin{figure}
\centering
\includegraphics[width=0.75\textwidth,height=\textheight]{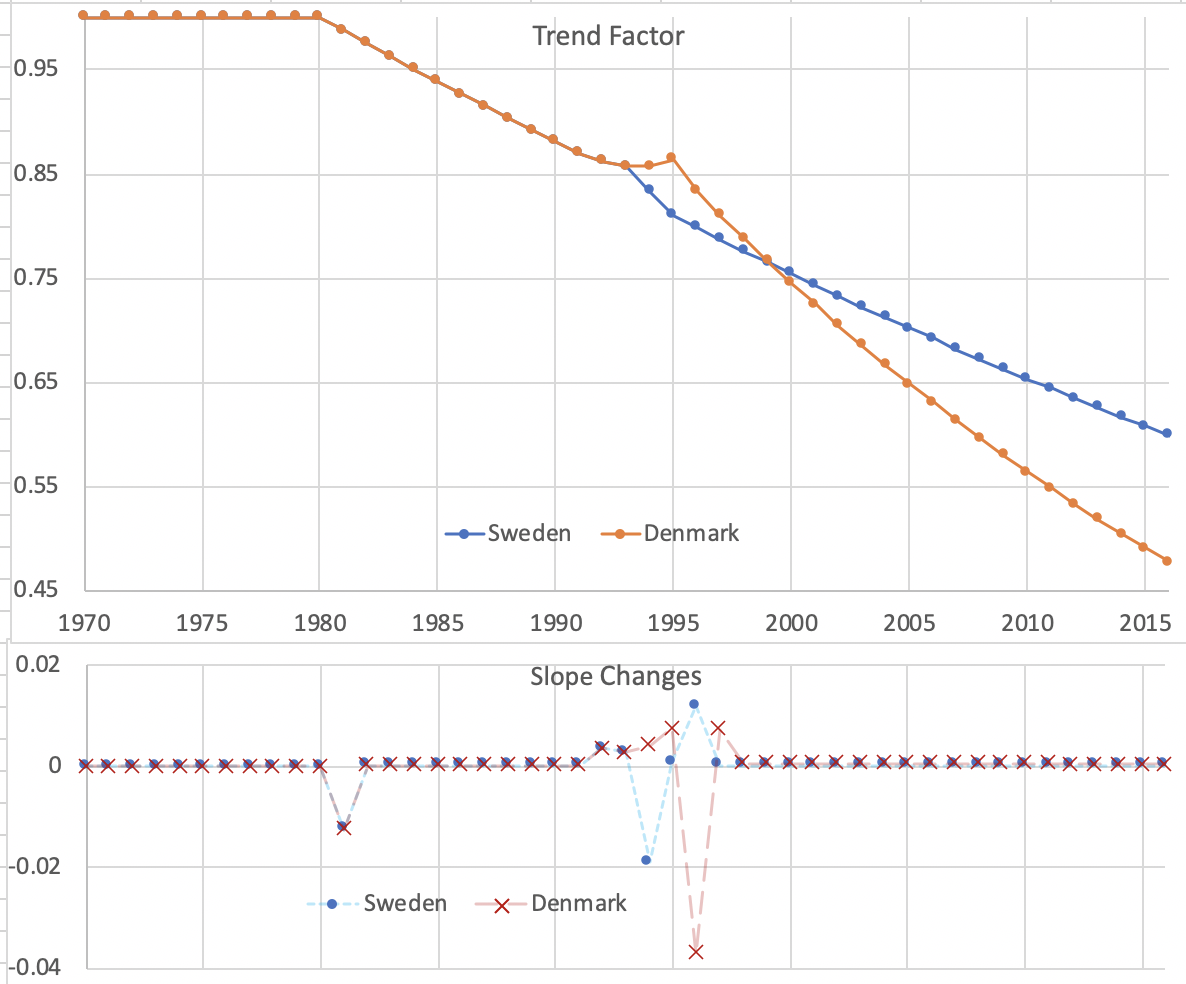}
\caption{Trend Factors\label{fig2}}
\end{figure}

\begin{figure}
\centering
\includegraphics[width=0.75\textwidth,height=\textheight]{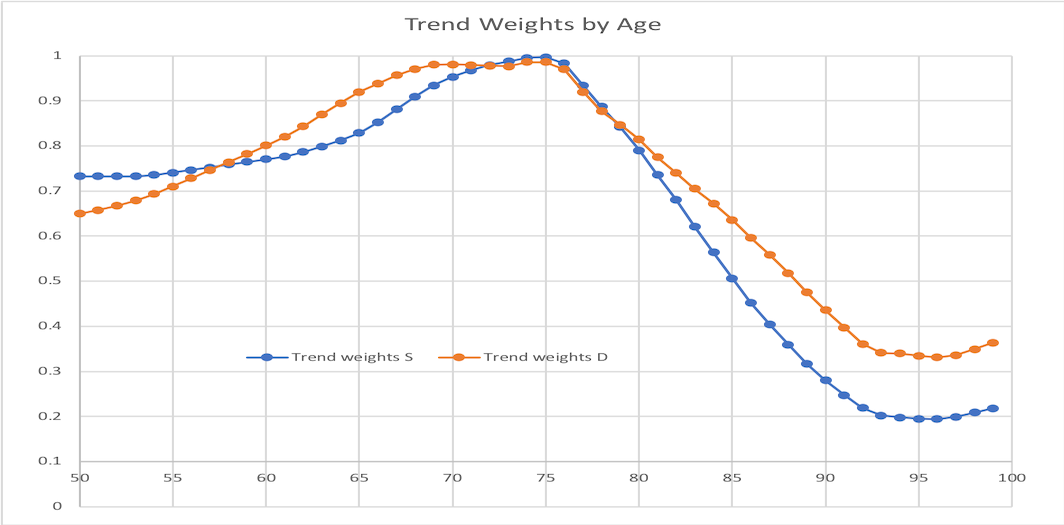}
\caption{Trend Age Weights\label{fig3}}
\end{figure}

\textbf{\emph{4b \textcolor{black}{ Renshaw-Haberman} model}}\\
The next step was to put in the trend weights by age, which gives the
\textcolor{black}{ Renshaw-Haberman} model. This is no longer a simple
linear model, which complicates the coding, but not substantially. We
used the constraint that the highest weight for any age for a population
would be 1.0, with all the weights positive. In that way, the trend at
any year is the trend for the age with the highest weight.

For this we needed a separate slope-change design matrix \(Y\) for trend
weights, with 100 rows and columns -- one for each age in each
population.
\textcolor{black}{$Y$ is to be multiplied by a column vector of 100 parameters which produce a vector of 50 trend weights for the first population and 50 for the second. For compactness, the variables with parameters close to zero were taken out, but this is not a necessary step and has minimal effect on the fitted values.}
When unneeded variables were eliminated, their columns were dropped but
not the rows. Then the parameter vector \(\eta\) times \(Y\) gives the
initial weights for each age.
\textcolor{black}{To make the weights positive with a maximum of 1.0, their maximum value is subtracted by population}
then exponentiated to give the age weights \(\alpha\).
\[\alpha_1 = Y\eta\]
\[\alpha = e^{\alpha_1-max(\alpha_1)}\]
\textcolor{black}{Now $\alpha$ has the trend weight for each age. These are multiplied by a $2ab$ x 100 0,1 design matrix with a 1 in each row, at the age and population column for that observation, giving $A$, a vector of the trend weights by observation. Next, the original design matrix columns are grouped into two parts:}
the design matrix \(X\) and the parameter vector \(\beta\) are only for
the age and cohort parameters,
\textcolor{black}{and the original columns for the years go into}
design matrix \(Z\) with parameters \(\psi.\)
\textcolor{black}{$Z\psi$ then is the trend for the right year for each observation. This is multiplied pointwise by the vector $A$ to give the weighted trend for each observation. }
Then \((i)\) becomes:
\[\hspace{13em}\widehat{m} = c+X\beta+ A\circ (Z\psi) \hspace{16em} (iii)\]\textcolor{black}{This is not strictly a linear model as parameters $\alpha$ and $\psi$ are combined multiplicatively.}
In the end, six more age, period, and cohort variables were eliminated,
leaving 46 APC variables, and 72 slope-change variables, so 118
variables all together. This produced a loo penalized NLL of 20,725.5,
an NLL of 20,644.5, which are improvements of about 1400 over the APC
model. The penalty of 81.0 for the 118 variables is a more typical
relationship for shrinkage models. NLL and penalized NL are in logs, so
it is the absolute difference, not the relative difference, than
indicates how much improvement a better model produces. Here that is
substantial, so Renshaw-Haberman is clearly better for this data.

\begin{figure}
\centering
\includegraphics[width=0.95\textwidth,height=\textheight]{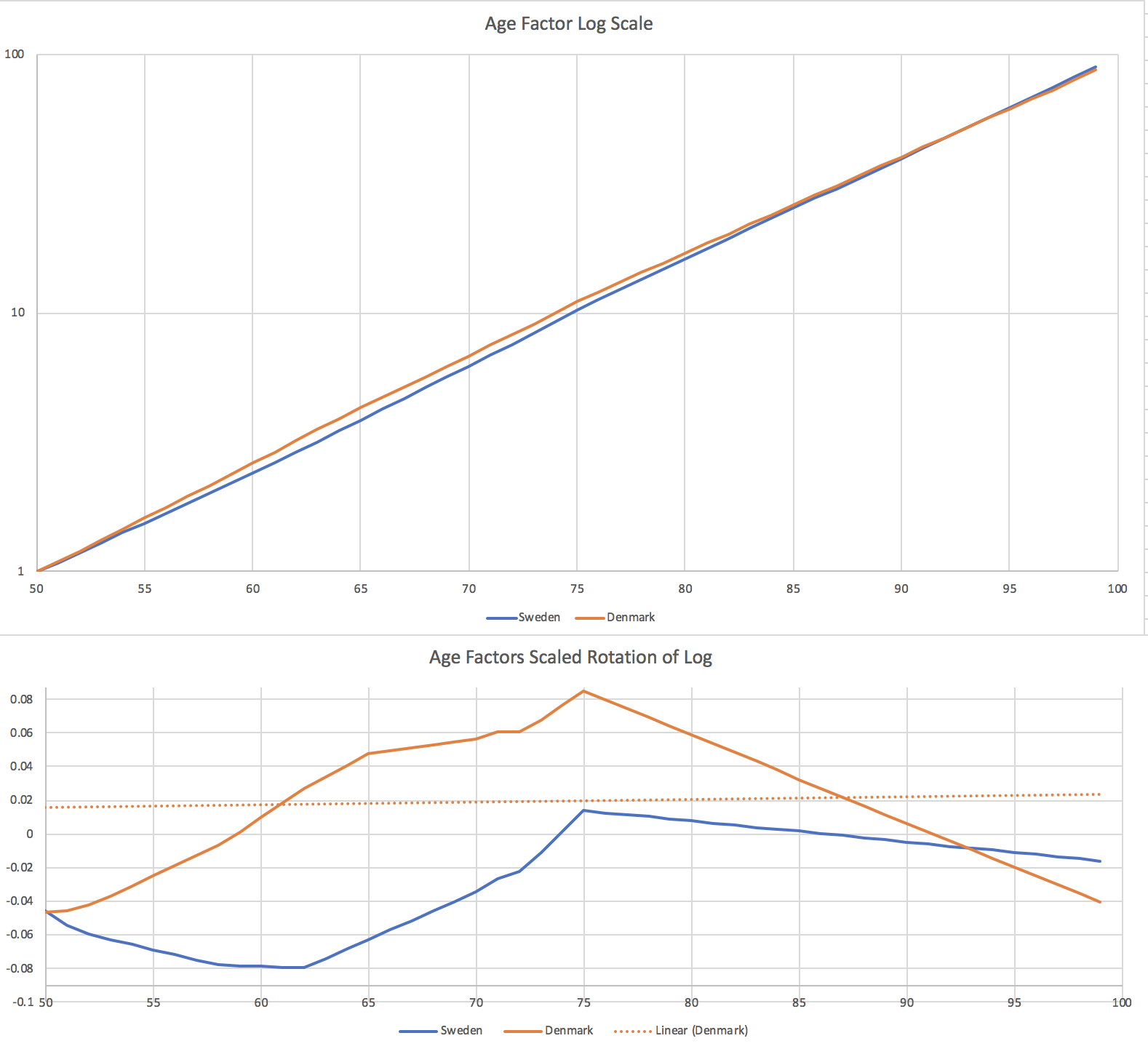}
\caption{Age Factors\label{fig4}}
\end{figure}

\textbf{\emph{4c Alternative shrinkage priors}}\\
Next we tried an alternative shrinkage distribution, as discussed in
Section 3, using the variables from the best model. Instead of the
double exponential priors, we assumed that the slope-change variables
were t-distributed. A double exponential is roughly equivalent to a t
with \(\nu = 6\), a Cauchy is \(\nu = 1\), and a normal is the limit as
\(\nu\) gets large. We tried a prior with \(log(\nu)\) uniform on
{[}-1,6{]}. That gives a range for \(\nu\) of about {[}0.37,403{]}. The
result was a slightly worse loo, at 20,726. The mean of \(\nu\) was 2.4,
with a median of 2. This is heavier tailed than the double exponential,
with larger parameters allowed and smaller ones pushed more towards
zero. Two more parameters came out \(< 0.001\) and can be eliminated.

We have found in other studies that loo does not change greatly with
small changes in the heaviness of the shrinkage distribution tails, and
it looked like estimating \(\nu\) was using up degrees of freedom. Next
we tried a constant \(\nu = 2\) and took out the two small trend-weight
variables. The result was then 116 variables, with a loo measure of
20,723.4, NLL of 20,643.9, and a penalty of 79.5. Such a drop of 2 in
the penalized NLL is usually considered a better model, and this one is
more parsimonious as well.

\begin{figure}
\centering
\includegraphics[width=0.75\textwidth,height=\textheight]{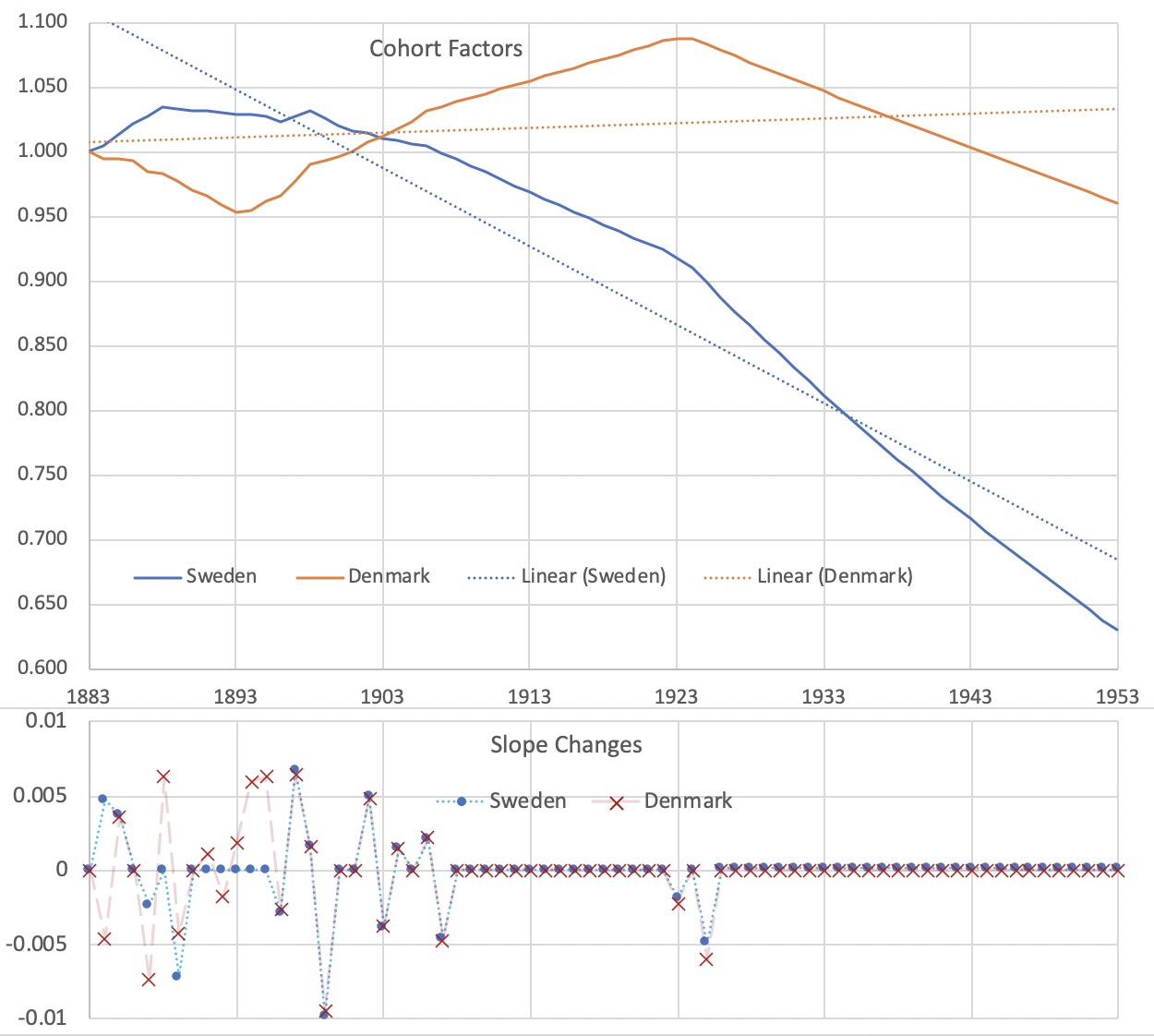}
\caption{Cohort Factors\label{fig5}}
\end{figure}

\textbf{\emph{4d Negative-binomial model}}\\
One assumption not checked so far is the Poisson distribution of deaths
in each age-period cell. As an alternative, we tried the negative
binomial distribution, which has an additional parameter \(\phi\), which
does not affect the mean but which makes the variance =
\(\mu^2/\phi + \mu\).
\textcolor{black}{This changes $(ii)$ to:$$\hspace{10em}d[j] \sim NB\left(\widetilde {e}[j]e^{\widehat{m}[j]},\phi\right)\hspace{10em} (iv)$$}This
actually gave a still lower NLL of 20,589.1, and loo of 20,648.1, so an
improvement of 75. The sample-bias penalty was much lower, at 59. There
were also three fewer trend-weight variables, leaving 67 of those, as
well as 46 age, period, and cohort variables as before. Now there is
still a constant, the shrinkage parameter, and \(\phi\) that are not
shrunk, so 116 variables in total. The low penalty means that the
loglikelihood of the omitted variables was not much worse than that of
the full model, which is part of why the predictive measure loo was so
much better.

The Poisson is an approximation to the binomial distribution that would
come from a sum of independent Bernoulli processes. Having a more
dispersed distribution suggests that there is some degree of correlation
in the deaths. That could arise from high deaths in bad weather or
disease outbreaks, for instance. However the dispersion is a measure of
the departure of the actual deaths from the fitted means, and also
includes model error and estimation error. Thus it is not entirely
certain that the better fit implies that deaths are correlated.

We just assumed that log(\(\phi\)) was distributed uniform on
\(\pm 1.7977\cdot10^{308}\). The conditional mean of \(\phi\) given the
data is 2839. The largest mean for any cell was slightly above 2000, and
that would get a variance increased by 70\%. This could be as little as
a 10\% increase for small cells. Compared to the Poisson, this makes it
less critical to get a very close fit on the largest cells, which can
give a better fit on the other cells. Of course, another reason for the
better fit could be actual greater dispersion. The parameters graphed in
Figures 3 -- 6 are for the negative-binomial model as are the actual
vs.~fitted values in Figures 7 and 8.

\begin{table}[]
\centering
\caption{Summary of Goodness-of-Fit Measures}
\begin{tabular}{lrrrr}
                & Parameters    & Loo & Penalty  & NLL \\
APC (Poisson with Laplace shrinkage)       & 54     & 22,159     & 77      & 22,082         \\
\textcolor{black}{ Renshaw-Haberman} (Poisson with Laplace shrinkage)   & 118   & 20,726     & 81      & 20,645            \\
\textcolor{black}{ Renshaw-Haberman} (Poisson with $\nu = 2$ shrinkage)  & 116 & 20,723 & 79 & 20,644     \\
\textcolor{black}{ Renshaw-Haberman} (Neg. Bin. with $\nu = 2$ shrinkage) & 116 & 20,648 & 59 & 20,589 \\    
\end{tabular}
\end{table}

\begin{table}[]
\centering
\caption{Parameter Count and Sum of Absolute Values for Negative Binomial Model}
\begin{tabular}{lrrrr}
                & Age    & Cohort & Trend  & Weight \\
Sweden Count       & 10     & 15     & 5      & 34         \\
Denmark Count      & 3      & 9      & 4      & 33         \\
Sweden Sum |$\beta$|   & 0.126 & 0.062 & 0.054 & 0.363     \\
Denmark Sum |$\beta$| & 0.022 & 0.041 & 0.093 & 0.189    \\ 
\end{tabular}
\end{table}

\begin{figure}
\centering
\includegraphics[width=0.9\textwidth,height=\textheight]{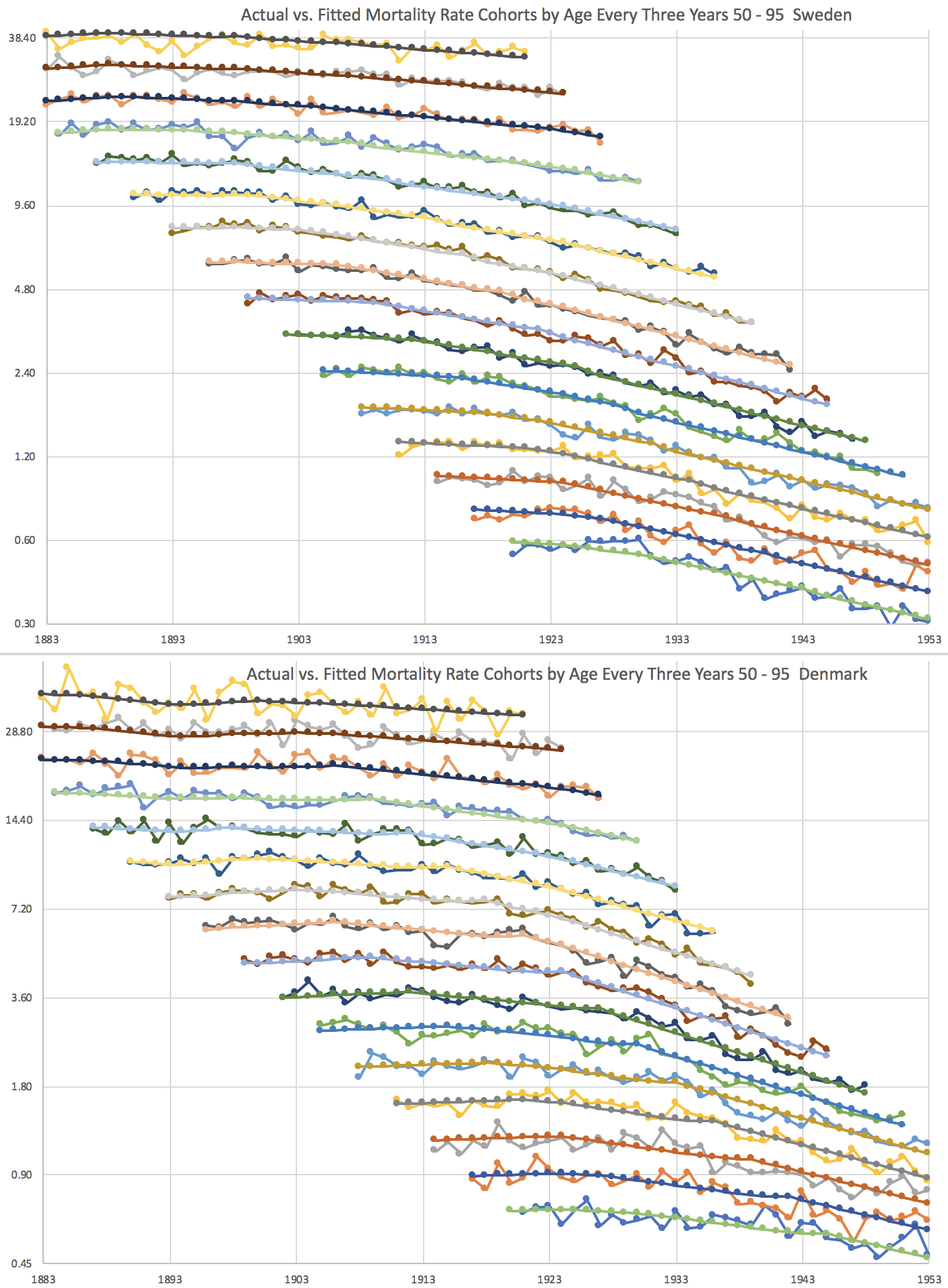}
\caption{Actual vs.~Fitted Age by Cohort\label{fig6}}
\end{figure}

\textbf{\emph{4e Fit summaries}}\\
Table 2 summarizes the goodness-of-fit measures for all the models.
Table 3 shows the number of parameters of each type by country and the
sum of their absolute values in the negative-binomial model. As we were
hoping, by doing the joint fitting, Denmark got fewer parameters and
lower impact as measured by parameter absolute values. Thus not so much
change from the Sweden model was needed. Trend is a bit of an exception,
and as we saw, Sweden does not have much trend while Denmark has some.
The mortality-by-age parameters are one area in particular where the
populations are similar, and Denmark does not need much change from the
Sweden values. It is surprising how dominant the age-weight parameters
are in both size and count, and also how they add so little to the NLL
penalty and so much to the goodness of fit. They are clearly important
in this model.

Figure 3 shows the parametric curves fit to trend factors by population.
Sweden has less trend in the later years. Denmark does not deviate from
the Swedish trend until halfway through the period. Both show long
stretches with few slope changes. These are a lot simpler than the
empirical trend graph in Figure 1, which could be why the age weights
are important. Figure 4 shows the curves for the age weights. They are
more complex than the trends, but still fairly smooth graphs. Denmark
largely has the same overall pattern as Sweden, so is using its
parameters to a fair degree. Figure 5 graphs the fitted age factors on a
log scale, along with a rotation of the factors to horizontal, to better
show their differences. There are some nuances of slope-change variation
that shows up in the rotation.

\begin{figure}
\centering
\includegraphics[width=0.9\textwidth,height=\textheight]{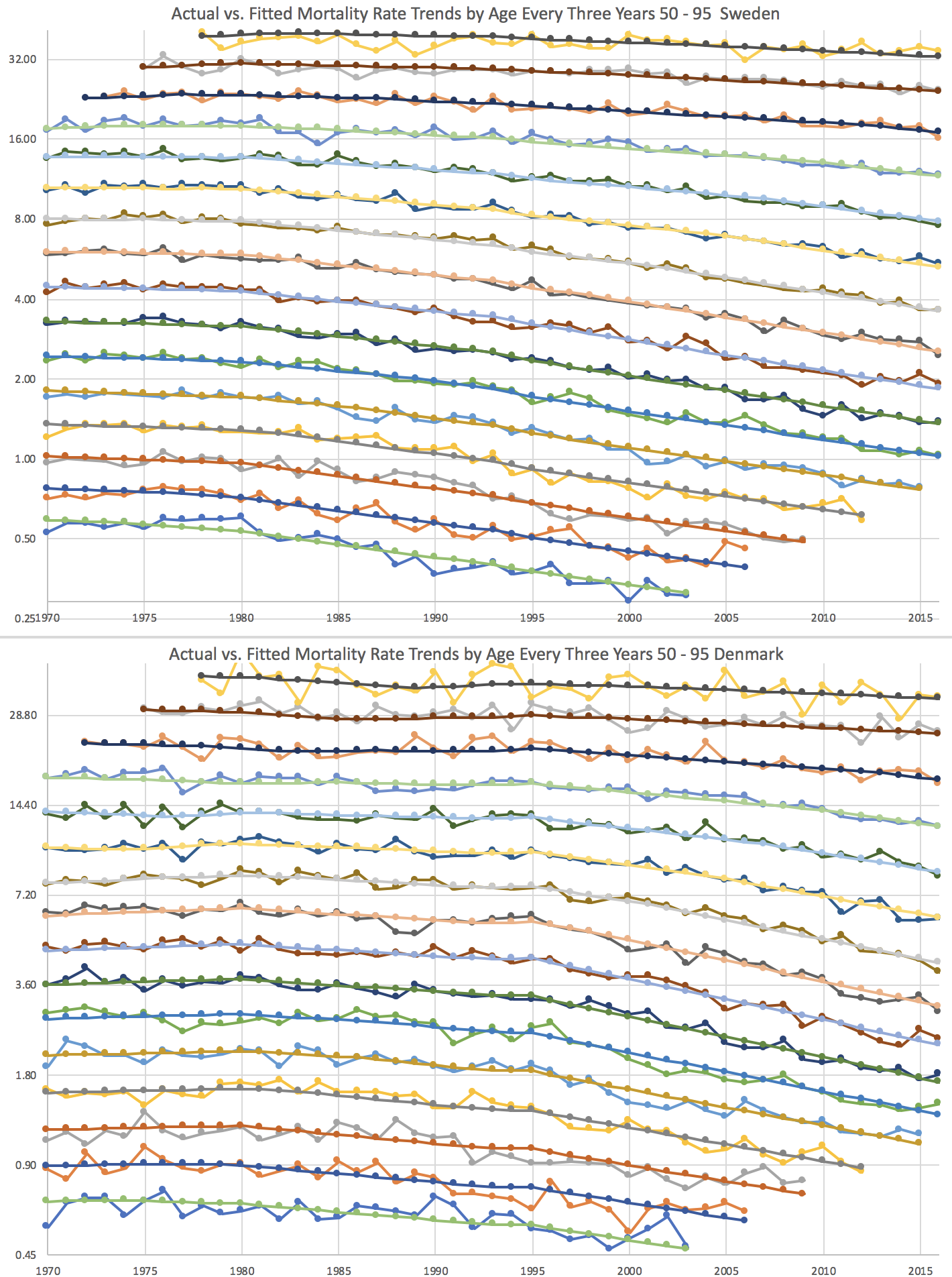}
\caption{Actual vs.~Fitted Age by Year\label{fig7}}
\end{figure}

Figure 6 shows the cohort factors, along with fitted lines to each to
show their average slopes. Although the shapes are different, a lot of
the slope changes and differences from linear are actually similar
between the two populations.

Actual vs.~fitted mortality rates are graphed for the cohorts by age in
3-year ranges in Figure 7. The fitted values form smooth curves in this
modeling. Denmark's rates are more volatile, but the fitted values
reasonably represent the trends in each country. For the most part, the
rates are lower for later cohorts. The cohort parameters did not do
exactly that, but the fitted rates are a combination of age, period,
cohort, and trend-weight parameters, which are all relative to each
other. Figure 8 graphs this for year of death instead of cohort, and the
conclusions are similar.

In the example, the difference parameters for the second population are
smaller and fewer than the parameters in the first model, suggesting
that it is retaining features of the model of the larger population. For
this data, the \textcolor{black}{ Renshaw-Haberman} model fits better
than the linear APC model, the negative binomial fits better than the
Poisson, and using the t-distribution with \(\nu = 2\) works well as an
assumed shrinkage distribution. A more parsimonious model, that looks
not as good by loo, can be obtained by a Cauchy shrinkage distribution
(the t-distribution with \(\nu = 1\)), which could be a good idea if the
model is regarded as an approximation to a more complex process. These
shrinkage distributions also provide minimization formulas similar to
those of lasso and ridge regression. The t with \(\nu = 2\) case, for
instance, leads to minimization of
NLL\(+1.5\sum _j log(2+\lambda^2\beta_j^2)\).

\textbf{\emph{\textcolor{black}{4f Projections}}}\\
\textcolor{black}{Projecting forward using these models would require projecting the period trends, and if more cohorts are to be included, the cohort trends would need to be projected as well. With the fits that we have, the period and cohort trends look remarkably stable for both countries. For over 20 years, the period trend for Sweden has been constantly downward, at a factor of 0.991 per year. For Denmark the trend rate has been 0.987 over this period. These are the factors for ages 70--75, with less trend at other ages according to the curves fit to trend weight by age. A reasonable starting point would be to continue these trend rates. You could not expect declining mortality forever, but the rates are not very steep. After 50 years of decline at this rate, Sweden's mortality at those ages would drop by a factor of 0.65, and Denmark's by 0.51. Other ages would have less decline, and the deaths overall would shift to still older ages.}

\textcolor{black}{The cohort trends are less steep, with factors of 0.992 and 0.996 for Sweden and Denmark, respectively. Fifty years of compounding of these would give reduction factors of 0.69 for Sweden and 0.81 for Denmark. If these were to be combined, those born in 2003 would have lower mortality by 0.43 in Sweden and 0.41 in Denmark for the peak trend ages. Actuaries would naturally want to temper these trends with reasonable judgement if they were to use them in annuity and insurance pricing. In practice, pandemic issues would of course now go into their calculations. These countries have data through the Spanish flu pandemic that might influence their adjustments.}

\textcolor{black}{Stan can output the parameters for every sampled posterior point. The prior distributions can be used to simulate future outcomes for each sample, and these could be used for distributions around the projected mean results. These priors all have mean zero, which would be consistent with continuing the latest trends. The correlations across the parameters could also be computed over the historical period and applied during the simulations.}

\textcolor{black}{To use a model like this in practice, it would be important to review the parameter distributions. The Appendix has some R code for running these models. That includes code for printing and graphing the parameter distributions.}

\textbf{\emph{\textcolor{black}{4g Population behavioral studies}}}\\
\textcolor{black}{Despite being among the countries with the longest life expectancy in the world from 1950 to 1980, Denmark experienced a dramatic fall in the rankings in the 1990s while Sweden maintained its position near the top. The difference in mortality between Denmark and Sweden has been discussed by several papers and}
Christensen et al. (2010)
\textcolor{black}{provides a comprehensive review as well as describing the trends in overall mortality and cause-specific mortality with a discussion based on the underlying determinants of reduced life expectancy in Denmark. They show that Denmark's life expectancy improvement rates were close to zero between the late 1980s and early 1990s. Although the mid-1990s, Denmark experienced an annual increase in life expectancy, Danish longevity has not been able to catch up with Sweden.}
Juel, Sorensen, and Bronnum-Hansen (2008)
\textcolor{black} {shows the significant impact of the risk factors such as alcohol consumption, smoking, physical inactivity and unhealthy diet on Danish life expectancy.}
Christensen et al. (2010)
\textcolor{black}{emphasizes the impact of smoking as the major explanation for the divergent Danish life expectancy trend compared to Sweden while acknowledging the contribution of different lifestyles and health care systems (investment in health care is lower in Denmark than in Sweden) in two countries. They also explain the reasons for the improvement in life expectancy in the early 1990s in Denmark as the effect of decreasing cardiovascular mortality, better lifestyles, better medical and surgical treatment as well as better medical disease prevention services.}

\textcolor{black} {Looking at the fitted period trends in Figure 3 and the fitted cohort trends in Figure 6 shows that our model estimates the period trends as being pretty similar until the mid-1990s, with Denmark improving faster after that. However the cohort trends are more favorable in Sweden. You could largely get the same overall effects by rotating the Swedish cohort curve up to match Denmark's and rotate its period curve down. Either would correspond to the empirical relationship between the countries. But the actual results from the estimation are those that best match the conditional distribution of the parameters given the data that come with no constraints, as discussed at the end of Section 3. Both populations' fits show sharp improvements beginning with cohorts born in the mid-1920s (who reached adulthood after WWII). This turn was slightly more in Denmark, but trends there had been upward before then, so Sweden maintains a faster rate of cohort improvement. Smoking was probably a key part of this. Denmark's slowdown in mortality improvement in the 1990s corresponds to when their worst cohorts by our fits reached their 70s. The cohort fits suggest that the unhealthy behaviors found in the studies noted above were largely a cohort effect. That is something that the lifestyle measures may be able to test.}

\textbf{5 Conclusion}\\
Joint APC-modeling of related populations is a common need in actuarial
work, including for mortality models and loss reserving. Semiparametric
regression is a fairly new approach for actuaries, and has advantages
over MLE in producing lower estimation and prediction errors. We show
how to apply it to individual mortality models by shrinking slope
changes of linear-spline fits to the parameters, and then to modeling
related data sets by shrinking changes in the semiparametric curves
between the data sets.

MCMC estimation can be applied to complex modeling problems. We show how
it can be used here to produce samples of the conditional distribution
of the parameters given the data. This is typically done in a nominally
Bayesian setting, where the model starts with postulated distributions
of the parameters, called priors, and yields the conditional
distribution of parameters given the data, called the posterior. But
frequentist random-effects modeling postulates distributions of the
effects (which look like parameters to Bayesians) and there seems to be
no reason it could not also use MCMC to sample from the conditional
distribution of the effects given the data. Furthermore, when Bayesians
take this approach, they often feel free to depart from the traditional
view of priors as incoming beliefs and treat them instead more like
random effects -- that is postulated distributions. Thus the approach we
take appears largely consistent with Bayesian and frequentist
approaches, but differs slightly from the historical practices of each.

MCMC software can have problems with highly correlated variables like
the slope change variables. Ridge regression and lasso were designed for
this problem, and can be much more efficient in estimating conditional
modes. We use lasso, with only minor shrinkage, to eliminate the
less-necessary slope-change variables, and then use MCMC with the
reduced parameter set. Common lasso software requires linear models, so
we start off with a lasso Poisson regression for the linear APC model,
with a log link, to reduce the variable set going into MCMC.
\textcolor{black}{We also present some advantages that MCMC shrinkage of linear-spline slope changes has over the standard semiparametric use of O'Sullivan penalized splines.}

\newpage

\textbf{\textcolor{black}{Appendix -- Code Examples}}\\
First we look at doing a Poisson lasso run in R starting with a
shrinkage design matrix. The cv.glmnet code exponentiates a regression
model to give the Poisson means. If an offset is included, effectively
the means before exponentiating are the offsets plus the regression fit,
and have an offset for every observation. Thus the offsets should be the
logs of the exposures. The constant term is not in the design matrix but
is added in.

\begin{Shaded}
\begin{Highlighting}[]
\KeywordTok{library}\NormalTok{(readxl)}
\KeywordTok{library}\NormalTok{(glmnet)}
\NormalTok{y =}\StringTok{ }\KeywordTok{as.integer}\NormalTok{(}\KeywordTok{scan}\NormalTok{(}\StringTok{'deaths.txt'}\NormalTok{))  }\CommentTok{#scan turns a column file into a vector}
\NormalTok{x =}\StringTok{ }\KeywordTok{as.matrix}\NormalTok{(}\KeywordTok{read_excel}\NormalTok{(}\StringTok{"shrink_design_mat.xlsx"}\NormalTok{)) }\CommentTok{#spreadsheet has header row}
\NormalTok{offs =}\StringTok{ }\KeywordTok{scan}\NormalTok{(}\StringTok{'logexpo.txt'}\NormalTok{)}
\NormalTok{fit =}\StringTok{ }\KeywordTok{cv.glmnet}\NormalTok{(x, y, }\DataTypeTok{standardize =} \OtherTok{FALSE}\NormalTok{, }\DataTypeTok{family =} \StringTok{"poisson"}\NormalTok{, }\DataTypeTok{offset=}\NormalTok{offs)}
\NormalTok{out <-}\StringTok{ }\KeywordTok{as.matrix}\NormalTok{(}\KeywordTok{coef}\NormalTok{(fit, }\DataTypeTok{s =}\NormalTok{ fit}\OperatorTok{$}\NormalTok{lambda.min)) }\CommentTok{#these are the lasso parameters}
\KeywordTok{write.csv}\NormalTok{(out, }\DataTypeTok{file=}\StringTok{"lasso_fit.csv"}\NormalTok{)}
\end{Highlighting}
\end{Shaded}

This is possible Stan code mort\_apc.stan for the APC case. The double
exponential = Laplace prior is known as Bayesian lasso as the posterior
mode is closely related to the lasso estimates.

\begin{Shaded}
\begin{Highlighting}[]
\NormalTok{data \{}
\NormalTok{  int N;     }\OperatorTok{/}\ErrorTok{/}\StringTok{ }\NormalTok{number of obs}
\NormalTok{  int U;    }\OperatorTok{/}\ErrorTok{/}\StringTok{ }\NormalTok{number of variables}
\NormalTok{  vector[N] expo;}
\NormalTok{  int y[N];}
\NormalTok{  matrix[N,U] x;}
\NormalTok{\}}
\NormalTok{parameters \{  }\OperatorTok{/}\ErrorTok{/}\StringTok{ }\NormalTok{all except v will get uniform prior, which is default         }
\NormalTok{ real}\OperatorTok{<}\NormalTok{lower=}\OperatorTok{-}\DecValTok{8}\NormalTok{, upper=}\OperatorTok{-}\DecValTok{3}\OperatorTok{>}\StringTok{ }\NormalTok{cn; }\OperatorTok{/}\ErrorTok{/}\NormalTok{constant term, starting }\ControlFlowTok{in}\NormalTok{ known range}
\NormalTok{ real}\OperatorTok{<}\NormalTok{lower=}\OperatorTok{-}\DecValTok{6}\NormalTok{, upper =}\StringTok{ }\DecValTok{-3}\OperatorTok{>}\StringTok{ }\NormalTok{logs; }\OperatorTok{/}\ErrorTok{/}\NormalTok{log of s, related to lambda, not too high}
\NormalTok{ vector[U] v;}
\NormalTok{\} }
\NormalTok{transformed parameters \{      }
\NormalTok{ real s;           }\OperatorTok{/}\ErrorTok{/}\StringTok{ }\NormalTok{shrinkage parameter}
\NormalTok{ vector[N] mu;}
\NormalTok{ s =}\StringTok{ }\KeywordTok{exp}\NormalTok{(logs); }\OperatorTok{/}\ErrorTok{/}\NormalTok{makes }\DecValTok{1}\OperatorTok{/}\NormalTok{x prior }\ControlFlowTok{for}\NormalTok{ s}\OperatorTok{>}\DecValTok{0}\NormalTok{ to prevent bias}
\NormalTok{ mu =}\StringTok{ }\KeywordTok{exp}\NormalTok{(cn}\OperatorTok{+}\NormalTok{x}\OperatorTok{*}\NormalTok{v);}
 \ControlFlowTok{for}\NormalTok{ (j }\ControlFlowTok{in} \DecValTok{1}\OperatorTok{:}\NormalTok{N) mu[j] =}\StringTok{ }\NormalTok{mu[j]}\OperatorTok{*}\NormalTok{expo[j];}
\NormalTok{\}}
\NormalTok{model \{  }\OperatorTok{/}\ErrorTok{/}\StringTok{ }\NormalTok{gives priors }\ControlFlowTok{for}\NormalTok{ those not assumed uniform. Choose this one }\ControlFlowTok{for}\NormalTok{ lasso.}
   \ControlFlowTok{for}\NormalTok{ (i }\ControlFlowTok{in} \DecValTok{1}\OperatorTok{:}\NormalTok{U)  v[i] }\OperatorTok{~}\StringTok{ }\KeywordTok{double_exponential}\NormalTok{(}\DecValTok{0}\NormalTok{, s);  }\OperatorTok{/}\ErrorTok{/}\StringTok{ }\NormalTok{more weight to close to }\DecValTok{0}
\NormalTok{y }\OperatorTok{~}\StringTok{ }\KeywordTok{poisson}\NormalTok{(mu);}
\NormalTok{\} }
\NormalTok{generated quantities \{ }\OperatorTok{/}\ErrorTok{/}\NormalTok{outputs log likelihood }\ControlFlowTok{for}\NormalTok{ loo}
\NormalTok{  vector[N] log_lik;}
 \ControlFlowTok{for}\NormalTok{ (j }\ControlFlowTok{in} \DecValTok{1}\OperatorTok{:}\NormalTok{N) log_lik[j] =}\StringTok{ }\KeywordTok{poisson_lpmf}\NormalTok{(y[j] }\OperatorTok{|}\StringTok{ }\NormalTok{mu[j]);}
\NormalTok{\}}
\end{Highlighting}
\end{Shaded}

This is R code for running the Stan code mort\_apc.stan and doing some
analysis. The print command allows you to print out distribution ranges
for each parameter with selected parameters and percentiles. Plot with
show density plots all of the selected parameters as density graphs on a
single scale. Plot with ``hist'' selected graphs histograms of each
parameter's posterior distribution. Extract gets all of the parameters
by sample, but you have to be careful to check what order they come out
in. One way to do this is just to keep the first chain, which here is
mort\_ss{[}\emph{,1,}{]}, and look at it as an array, which will have
variable names as column headings. The order of the variables will be
the same if you do it again with all of the chains. With the parameter
sample distributions you can also compute the correlations among the
parameters. These could be used along with the prior distributions to
simulate parameter changes going forward to get projections ranges.

\begin{Shaded}
\begin{Highlighting}[]
\KeywordTok{library}\NormalTok{(readxl)  }\CommentTok{# allows reading excel files; assumes there is a header row}
\KeywordTok{library}\NormalTok{(}\StringTok{"loo"}\NormalTok{)}
\KeywordTok{library}\NormalTok{(rstan)}
\KeywordTok{rstan_options}\NormalTok{(}\DataTypeTok{auto_write =} \OtherTok{TRUE}\NormalTok{)}
\KeywordTok{options}\NormalTok{(}\DataTypeTok{mc.cores =}\NormalTok{ parallel}\OperatorTok{::}\KeywordTok{detectCores}\NormalTok{())}
 \CommentTok{#read in data}
\NormalTok{x =}\StringTok{ }\KeywordTok{as.matrix}\NormalTok{(}\KeywordTok{read_excel}\NormalTok{(}\StringTok{'shrink_design_mat.xlsx'}\NormalTok{))}
\NormalTok{y =}\StringTok{ }\KeywordTok{as.integer}\NormalTok{(}\KeywordTok{scan}\NormalTok{(}\StringTok{'deaths.txt'}\NormalTok{))}
\NormalTok{expo =}\StringTok{ }\KeywordTok{as.integer}\NormalTok{(}\KeywordTok{scan}\NormalTok{(}\StringTok{'expo.txt'}\NormalTok{))}
\NormalTok{N <-}\StringTok{ }\KeywordTok{nrow}\NormalTok{(x)}
\NormalTok{U <-}\StringTok{ }\KeywordTok{ncol}\NormalTok{(x)}
\KeywordTok{c}\NormalTok{(N,U)}
\NormalTok{df =}\StringTok{ }\KeywordTok{list}\NormalTok{(}\DataTypeTok{N=}\NormalTok{N,}\DataTypeTok{U=}\NormalTok{U,}\DataTypeTok{expo=}\NormalTok{expo,}\DataTypeTok{y=}\NormalTok{y,}\DataTypeTok{x=}\NormalTok{x)}
 \CommentTok{#now run stan}
\KeywordTok{set.seed}\NormalTok{(}\DecValTok{8}\NormalTok{)}
\NormalTok{mort_}\DecValTok{1}\NormalTok{ <-}\StringTok{ }\KeywordTok{stan}\NormalTok{(}\DataTypeTok{file =} \StringTok{'mort_apc.stan'}\NormalTok{, }\DataTypeTok{data=}\NormalTok{df, }\DataTypeTok{verbose =} \OtherTok{FALSE}\NormalTok{, }\DataTypeTok{chains =} \DecValTok{4}\NormalTok{, }
        \DataTypeTok{iter =} \DecValTok{2000}\NormalTok{, }\DataTypeTok{control =} \KeywordTok{list}\NormalTok{(}\DataTypeTok{adapt_delta =} \FloatTok{0.9}\NormalTok{, }\DataTypeTok{max_treedepth =} \DecValTok{14}\NormalTok{))}
 \CommentTok{#compute loo}
\NormalTok{log_spread_}\DecValTok{1}\NormalTok{ <-}\StringTok{ }\KeywordTok{extract_log_lik}\NormalTok{(mort_}\DecValTok{1}\NormalTok{) }
\NormalTok{loo_spread_}\DecValTok{1}\NormalTok{ <-}\StringTok{ }\KeywordTok{loo}\NormalTok{(log_spread_}\DecValTok{1}\NormalTok{)}
\NormalTok{loo_spread_}\DecValTok{1}
 \CommentTok{#output parameter means by chain}
\NormalTok{out_mort_}\DecValTok{1}\NormalTok{ <-}\StringTok{ }\KeywordTok{get_posterior_mean}\NormalTok{(mort_}\DecValTok{1}\NormalTok{)}
\KeywordTok{write.csv}\NormalTok{(out_mort_}\DecValTok{1}\NormalTok{, }\DataTypeTok{file=}\StringTok{"out_apc.csv"}\NormalTok{)}
 \CommentTok{#show some output}
\KeywordTok{plot}\NormalTok{(mort_}\DecValTok{1}\NormalTok{, }\DataTypeTok{pars =} \StringTok{"v"}\NormalTok{, }\DataTypeTok{show_density =} \OtherTok{TRUE}\NormalTok{, }\DataTypeTok{ci_level =} \FloatTok{0.8}\NormalTok{, }\DataTypeTok{fill_color =} \StringTok{"black"}\NormalTok{)}
\KeywordTok{plot}\NormalTok{(mort_}\DecValTok{1}\NormalTok{, }\DataTypeTok{plotfun =} \StringTok{"hist"}\NormalTok{, }\DataTypeTok{pars =} \StringTok{"v"}\NormalTok{) }\CommentTok{#histograms for each parameter}
\KeywordTok{print}\NormalTok{(mort_}\DecValTok{1}\NormalTok{, }\DataTypeTok{pars=}\KeywordTok{c}\NormalTok{(}\StringTok{"c"}\NormalTok{, }\StringTok{"v"}\NormalTok{, }\StringTok{"s"}\NormalTok{), }\DataTypeTok{probs=}\KeywordTok{c}\NormalTok{(.}\DecValTok{025}\NormalTok{, }\FloatTok{0.2}\NormalTok{, }\FloatTok{0.5}\NormalTok{, }\FloatTok{0.8}\NormalTok{, }\FloatTok{0.975}\NormalTok{), }
      \DataTypeTok{digits_summary =} \DecValTok{5}\NormalTok{)}
 \CommentTok{#get all samples}
\NormalTok{mort_ss =}\StringTok{ }\KeywordTok{extract}\NormalTok{(mort_}\DecValTok{1}\NormalTok{, }\DataTypeTok{permuted =} \OtherTok{FALSE}\NormalTok{) }
\CommentTok{#3D array iterations by chains by parameters}
\CommentTok{#can get rid of unneeded columns, concatenate to 2 dimensions, write out, etc.}
\CommentTok{#array variables can be in different order than from get_posterior_mean}
\CommentTok{#if you just keep 1 chain you can write out as .csv and it will have headings}
\end{Highlighting}
\end{Shaded}

\textbf{References}

\hypertarget{refs}{}
\leavevmode\hypertarget{ref-antonio2015}{}%
Antonio, K., A. Bardoutsos, and W. Ouburg. 2015. ``Bayesian Poisson
Log-Bilinear Models for Mortality Projections with Multiple
Populations.'' \emph{European Actuarial Journal} 5: 245--81.

\leavevmode\hypertarget{ref-barnett2000}{}%
Barnett, Glen, and Ben Zehnwirth. 2000. ``Best Estimates for Reserves.''
\emph{Proceedings of the Casualty Actuarial Society} 87: 245--303.

\leavevmode\hypertarget{ref-buhlmann1967}{}%
Bühlmann, Hans. 1967. ``Experience Rating and Credibility.'' \emph{Astin
Bulletin} 4:3: 199--207.

\leavevmode\hypertarget{ref-cairns2011}{}%
Cairns, A. J. G., D. Blake, K. Dowd, G. D. Coughlan, and M.
Khalaf-Allah. 2011. ``Bayesian Stochastic Mortality Modelling for Two
Populations.'' \emph{Astin Bulletin} 41:1: 29--59.

\leavevmode\hypertarget{ref-yang2018}{}%
Chang, Yang, and Michael Sherris. 2018. ``Longevity Risk Management and
the Development of a Value-Based Longevity Index.'' \emph{Risks, MDPI,
Open Access Journal} 6(1) February: 1--20.

\leavevmode\hypertarget{ref-chris2010}{}%
Christensen, Kaare, Michael Davidsen, Knud Juel, Laust Mortensen, Roland
Rau, and James W Vaupel. 2010. ``The Divergent Life-Expectancy Trends in
Denmark and Sweden---and Some Potential Explanations.'' \emph{In
International Differences in Mortality at Older Ages: Dimensons and
Sources., Edited by E M Crimmins, S H Preston, and B Cohen}, 385--408.
\url{https://www.ncbi.nlm.nih.gov/books/NBK62592/pdf/Bookshelf_NBK62592.pdf}.

\leavevmode\hypertarget{ref-craven1979}{}%
Craven, P., and G. Wahba. 1979. ``Smoothing Noisy Data with Spline
Functions: Estimating the Correct Degree of Smoothing by the Method of
Generalized Cross-Validation.'' \emph{Numerische Mathematik} 31:
377--403.

\leavevmode\hypertarget{ref-dowd2011}{}%
Dowd, K., A. J. G. Cairns, D. Blake, G. D. Coughlan, D. Epstein, and M.
Khalaf-Allah. 2011. ``A Gravity Model of Mortality Rates for Two Related
Populations.'' \emph{North American Actuarial Journal} 15:2: 334--56.

\leavevmode\hypertarget{ref-friedman2010}{}%
Friedman, Jerome, Trevor Hastie, and Robert Tibshirani. 2010.
``Regularization Paths for Generalized Linear Models via Coordinate
Descent.'' \emph{Journal of Statistical Software} 33(1): 1--22.
\url{https://www.jstatsoft.org/article/view/v033i01}.

\leavevmode\hypertarget{ref-gao2018}{}%
Gao, Guangyuan, and S. Meng. 2018. ``Stochastic Claims Reserving via a
Bayesian Spline Model with Random Loss Ratio Effects.'' \emph{Astin
Bulletin} 48:1: 55--88.

\leavevmode\hypertarget{ref-gelfand1996}{}%
Gelfand, A. E. 1996. ``Model Determination Using Sampling-Based
Methods.'' \emph{Markov Chain Monte Carlo in Practice, Ed. W. R. Gilks,
S. Richardson, D. J. Spiegelhalter} London: Chapman and Hall: 145--62.

\leavevmode\hypertarget{ref-haberman2011}{}%
Haberman, S., and A. E. Renshaw. 2011. ``A Comparative Study of
Parametric Mortality Projection Models.'' \emph{Insurance: Mathematics
and Economics} 48 (1): 35--55.

\leavevmode\hypertarget{ref-harez2018}{}%
Harezlak, Jaroslaw, David Ruppert, and Matt P. Wand. 2018.
``Semiparametric Regression with R.'' \emph{Springer, NY}.

\leavevmode\hypertarget{ref-hastie2017}{}%
Hastie, Trevor, Robert Tibshirani, and Jerome Friedman. 2017. ``The
Elements of Statistical Learning.'' \emph{Springer} Corrected 12th
Printing.
\url{https://web.stanford.edu/~hastie/ElemStatLearn//printings/ESLII_print12.pdf}.

\leavevmode\hypertarget{ref-hoerl1970}{}%
Hoerl, A. E., and R. Kennard. 1970. ``Ridge Regression: Biased
Estimation for Nonorthogonal Problems.'' \emph{Technometrics} 12:
55--67.

\leavevmode\hypertarget{ref-hmd}{}%
Human Mortality Database. 2019. ``University of California, Berkeley
(Usa), Max Planck Institute for Demographic Research (Germany) and
United Nations.'' \url{https://www.mortality.org/}.

\leavevmode\hypertarget{ref-hunt2014}{}%
Hunt, Andrew, and David Blake. 2014. ``A General Procedure for
Constructing Mortality Models.'' \emph{North American Actuarial Journal}
18 (1): 116--38.

\leavevmode\hypertarget{ref-jarner2009}{}%
Jarner, S. F., and E. M. Kryger. 2009. ``Modelling Adult Mortality in
Small Populations: The Saint Model.'' \emph{Pensions Institute}
Discussion Paper PI-0902.

\leavevmode\hypertarget{ref-juel2008}{}%
Juel, K., J. Sorensen, and H. Bronnum-Hansen. 2008. ``Risk Factors and
Public Health in Denmark.'' \emph{Scandinavian Journal of Public Health
(Supplement 1)} 36: 112--227.
\url{https://journals.sagepub.com/doi/pdf/10.1177/1403494800801101}.

\leavevmode\hypertarget{ref-lee1992}{}%
Lee, R., and L. Carter. 1992. ``Modeling and Forecasting U.S.
Mortality.'' \emph{Journal of the American Statistical Association} 87:
659--75.

\leavevmode\hypertarget{ref-li2005}{}%
Li, N., and R. Lee. 2005. ``Coherent Mortality Forecasts for a Group of
Populations: An Extension of the Lee-Carter Method.'' \emph{Demography}
42:3: 575--94.

\leavevmode\hypertarget{ref-mowbray1914}{}%
Mowbray, Albert H. 1914. ``How Extensive a Payroll Exposure Is Necessary
to Give a Dependable Pure Premium.'' \emph{Proceedings of the Casualty
Actuarial Society} 1: 24--30.

\leavevmode\hypertarget{ref-osull1986}{}%
O'Sullivan, Finbarr. 1986. ``A Statistical Perspective on Ill-Posed
Inverse Problems.'' \emph{Statistical Science 1:4}, 502:518.

\leavevmode\hypertarget{ref-renshaw2006}{}%
Renshaw, A. E., and S. Haberman. 2006. ``A Cohort-Based Extension to the
Lee-Carter Model for Mortality Reduction Factors.'' \emph{Insurance:
Mathematics and Economics} 38: 556--70.

\leavevmode\hypertarget{ref-santosa1986}{}%
Santosa, Fadil, and William W. Symes. 1986. ``Linear Inversion of
Band-Limited Reflection Seismograms.'' \emph{SIAM Journal on Scientific
and Statistical Computing} 7(4): 1307--30.

\leavevmode\hypertarget{ref-rstan}{}%
Stan Development Team. 2020. ``RStan: The R Interface to Stan.''
\url{http://mc-stan.org/}.

\leavevmode\hypertarget{ref-stein1956}{}%
Stein, Charles. 1956. ``Inadmissibility of the Usual Estimator of the
Mean of a Multivariate Normal Distribution.'' \emph{Proceedings of the
Third Berkeley Symposium} 1: 197--206.

\leavevmode\hypertarget{ref-tibs1996}{}%
Tibshirani, Robert. 1996. ``Regression Shrinkage and Selection via the
Lasso.'' \emph{Journal of the Royal Statistical Society. Series B
(Methodological)} 58(1): 267--88.

\leavevmode\hypertarget{ref-tikhonov1943}{}%
Tikhonov, Andrey Nikolayevich. 1943. ``On the Stability of Inverse
Problems.'' \emph{Doklady Akademii Nauk SSSR} 39:5: 195--98.

\leavevmode\hypertarget{ref-vehtari2017}{}%
Vehtari, Aki, Andrew Gelman, and Jonah Gabry. 2017. ``Practical Bayesian
Model Evaluation Using Leave-One-Out Cross-Validation and Waic.''
\emph{Journal of Statistics and Computing} 27:5: 1413--32.

\leavevmode\hypertarget{ref-venter2015}{}%
Venter, Gary, Roman Gutkovich, and Qian Gao. 2019. ``Parameter Reduction
in Actuarial Triangle Models.'' \emph{Variance} 12:2: 142--60.

\leavevmode\hypertarget{ref-venter2018}{}%
Venter, Gary, and Şule Şahin. 2018. ``Parsimonious Parameterization of
Age-Period-Cohort Models by Bayesian Shrinkage.'' \emph{Astin Bulletin}
48:1: 89--110.

\leavevmode\hypertarget{ref-xu2019}{}%
Xu, Yajing, Michael Sherris, and Jonathan Ziveyi. 2019.
``Continuous-Time Multi-Cohort Mortality Modelling with Affine
Processes.'' \emph{Scandinavian Actuarial Journal}, 1--27.

\end{document}